\documentclass[journal]{IEEEtran}
\usepackage{multirow,bigstrut}
\usepackage[normalem]{ulem}
\useunder{\uline}{\ul}{}
\usepackage{graphicx}
\usepackage{float}
\usepackage{subfigure}
\usepackage{amsmath,amssymb,mathrsfs,bm}
\usepackage{algorithmic}
\usepackage{array}
\usepackage{fixltx2e}
\usepackage{stfloats,float}
\usepackage{arydshln} 

\usepackage[nocompress]{cite}
\usepackage[colorlinks]{hyperref}
\usepackage{xurl}

\usepackage{tikz,pgfplots}

\usepackage{amsthm}

\newtheorem{theorem}{Theorem}
\newtheorem{definition}[theorem]{Definition}

\newcommand\mtensor[1]{$\tensor{#1}$}
\newcommand\tensor[1]{\mathcal{#1}}

\begin{document}
	
\title{Haar Nuclear Norms with Applications to Remote Sensing Imagery Restoration}

\author{Shuang Xu, 
    Chang Yu,
    Jiangjun Peng,
    Xiangyong Cao,
    and Deyu Meng 
\thanks{Shuang Xu and Jiangjun Peng are with the School of Mathematics and Statistics, Northwestern Polytechnical University, Xi'an 710021, China (e-mail: xs@nwpu.edu.cn; andrew.pengjj@gmail.com). Chang Yu and Deyu Meng are with the School of Mathematics and Statistics, Xi'an Jiaotong University, Xi'an 710049, China. (e-mail: yuchang@stu.xjtu.edu.cn; dymeng@mail.xjtu.edu.cn). Xiangyong Cao is with School of Electronic and Information Engineering and the Key Laboratory of Intelligent Networks and Network Security,
	Ministry of Education, Xi'an Jiaotong University, Xi'an 710049, China (e-mail: caoxiangyong@mail.xjtu.edu.cn).}
}


\maketitle

\begin{abstract}
Remote sensing image restoration aims to reconstruct missing or corrupted areas within images. To date, low-rank based models have garnered significant interest in this field. This paper proposes a novel low-rank regularization term, named the Haar nuclear norm (HNN), for efficient and effective remote sensing image restoration. 
	It leverages the low-rank properties of wavelet coefficients derived from the 2-D frontal slice-wise Haar discrete wavelet transform, effectively modeling the low-rank prior for separated coarse-grained structure and fine-grained textures in the image. Experimental evaluations conducted on hyperspectral image inpainting, multi-temporal image cloud removal, and hyperspectral image denoising have revealed the HNN's potential. Typically, HNN achieves a performance improvement of 1-4 dB and a speedup of 10-28x compared to some state-of-the-art methods (e.g., tensor correlated total variation, and fully-connected tensor network) for inpainting tasks.
\end{abstract}

\begin{IEEEkeywords}
tensor completion, hyperspectral image denoising, remote sensing image cloud removal.
\end{IEEEkeywords}

\IEEEpeerreviewmaketitle

\section{Introduction}
\IEEEPARstart{R}{emote} sensing imagery restoration aims to reconstruct missing or corrupted areas of images \cite{Hyperspectral_Review}, including hyperspectral images (HSIs) and multi-temporal images (MTIs), typically represented as multi-dimensional arrays. Inpainting and denoising are key tasks in this restoration process \cite{QiuZZX24,10144690}. The mathematical formulations for these tasks are as follows:
\begin{equation}
	\min_{\tensor{X}} R(\tensor{X}) \quad \text{s. t.} \quad \mathscr{P}_{\Omega}(\tensor{M})=\mathscr{P}_{\Omega}(\tensor{X}),
\end{equation}
and
\begin{equation}
	\min_{\tensor{X}} R(\tensor{X}) \quad \text{s. t.} \quad \tensor{M}=\tensor{X}+\tensor{E},
\end{equation}
Here $\tensor{E}$ represents noise, $\tensor{M}$ is the corrupted image, and $\tensor{X}$ is the recovered one. Here, $\Omega$ indicates the set of indices for observable pixels, $\mathscr{P}_{\Omega}(\cdot)$ is the orthogonal projection onto the observable pixels, and $R(\cdot)$ is the regularization term applied to the recovered images to ensure desired properties such as smoothness or low-rankness.

This problem is typically addressed by low-rank (LR) models, which exploits the intrinsic structure and correlations in the data, under the assumption that data lie in a low-dimensional subspace \cite{ZhuangN20,WangPCWZM23}. These images tend to exhibit spatial and channel redundancies, making LR models particularly effective. The nuclear norm (NN) and its variants, such as the tensor nuclear norm (TNN) \cite{lu2018exact,lu2019low}, are commonly used as LR regularization. TNN, for instance, is extensively applied in various tensor processing problems and is defined as the sum of singular values from the first frontal slice of the singular value tensor, derived from tensor singular value decomposition (TSVD) \cite{TSVD}. Moreover, adjacent pixels in images often have similar values, leading to gradient sparsity. This property gives rise to a new optimization problem:
\begin{equation}
	\min_{\tensor{X}} R(\tensor{X}) + \lambda \|\nabla \tensor{X}\|_{1}, \quad \text{s. t.} \quad \mathscr{P}_{\Omega}(\tensor{M})=\mathscr{P}_{\Omega}(\tensor{X}),
\end{equation}
where $\|\nabla \tensor{X}\|_{1}$ represents the anisotropic total variation (TV) \cite{LRTV}, which is the $\ell_{1}$ norm of the gradient map $\nabla \tensor{X}$, and $\lambda > 0$ is the tuning parameter that balances the regularization terms. TV improves image restoration but is sensitive to the $\lambda$ parameter selection, limiting its practicality.

The recently proposed correlated TV (CTV) \cite{CTV} and tensor CTV (TCTV) \cite{TCTV} address this issue by imposing NN and TNN on the gradient map. For instance, TCTV, defined as the sum of TNNs of gradients along each mode, inherits the LR property from the original tensor and promotes gradient sparsity. In essence, TCTV simultaneously encourages LR and smoothness priors without parameter tuning. Most significantly, minimizing TCTV can achieve exact data recovery with high probability. In contrast to NN and TNN, which models LR prior in the image domain where low- and high-frequencies are entangled, the efficacy of CTV and TCTV stems from its ability to model the high-frequency LR prior, specifically the low-rankness of gradient maps.  

However, TCTV and CTV face certain challenges. Firstly, its computation is time-consuming, particularly for high-order images with a significantly larger number of channels than RGB images. For instance, processing an HSI of size $256\times256\times193$ can take TCTV approximately 2000 seconds, which is 10 times longer than the processing time for TNN. Secondly, while TCTV and CTV effectively models the high-frequency LR prior, neglecting the low-frequency information may result in the loss of crucial data and the color/spectral distortion. This study aims to address these critical issues, and its main contributions are outlined as follows:

(1) Drawing inspiration from the Haar wavelet transform (HWT), we propose a two-dimensional frontal HWT tailored for high-order images to disentangle low- and high-frequencies. Furthermore, we introduce a novel regularizer, the Haar nuclear norm (HNN), which models the LR property for both disentangled low- and high-frequency information, thereby combining LR and smoothness priors at a reduced scale. 

(2) The paper establishes the theory of exact recovery for HNN-regularized inpainting and denoising models, providing theoretical guarantees for the effectiveness of the proposed method.

(3) The paper develops an algorithm based on the Alternating Direction Method of Multipliers (ADMM) for solving the HNN-based restoration models. This algorithm ensures efficient computation, making HNN practical for real-world applications. For example, in inpainting experiments, HNN achieves a 1-4dB improvement in performance over state-of-the-art methods with a speedup of 10-28x.

\section{Related work}
\subsection{Low-rank models}
Low-rank models can be broadly classified into two categories: matrix/tensor decompositions and low-rank regularization techniques.

\subsubsection{Low-rank decomposition models} 
Matrix decompositions, with SVD as a prime example, are fundamental in low-rank models. SVD represents a matrix as a product of three matrices, enabling it to be expressed as a sum of rank-one matrices, which is key to many low-rank methods. For high-order images, tensor decompositions extend this concept, providing a more suitable framework to capture complex data relationships. 

The Canonical Polyadic (CP) decomposition \cite{CPD} factorizes a tensor into a sum of rank-one tensors, similar to how SVD works for matrices. The Tucker decomposition \cite{TuckerD} offers a more flexible representation by expressing a tensor as a core tensor multiplied by factor matrices along each mode, capturing varying correlations. Tensor Train (TT) \cite{TT,KoBDYW20} and Tensor Ring (TR) \cite{TR,LongZLL21} decompositions represent tensors as chains or rings of third-order tensors, effectively completing images. The recently proposed Fully Connected Tensor Network (FCTN) \cite{FCTN,zheng2024svdinstn,zheng2024provable} decomposes a tensor into a sequence of small tensors of the same order, with these tensors interacting to model the relationships between every pair of modes. FCTN holds promise for superior reconstruction of missing pixels \cite{SST_FCTN,ZhengNLFCTN} but may require careful rank selection. Recently, an architecture searching strategy has enhanced FCTN's data adaptability \cite{zheng2024svdinstn}. Besides exploring advanced decomposition models, the noise modeling techniques, such as non-independent and identically distributed (non-i.i.d.) \cite{NMOG,BALMF} or asymmetric \cite{MOAL2,MoAL3} noise models, have shown promising results for hyperspectral image processing. 

\subsubsection{Low-rank regularizations}
LR regularization encourages solutions with a low-rank structure. For matrices, the NN, the sum of singular values, is a convex relaxation of rank and is used in matrices. Nonetheless, the application of LR regularization to tensors remains an unresolved issue. The Sum of Nuclear Norm (SNN) \cite{liu2012tensor} is the sum of the nuclear norms of all possible matrix representations (matricizations) of a tensor, aiming for low ranks in all two-dimensional unfoldings, thus implicitly encouraging a low-rank structure for the entire tensor. Building upon the recently proposed tensor-tensor product, the TNN \cite{lu2018exact,lu2019low} is derived and demonstrated to be the convex envelope of the tensor average rank within the unit ball of the tensor spectral norm. In contrast to SNN, TNN provides theoretical guarantees of recovery. Recently, some more advanced norms for tensors are developed for denoising and inpainting \cite{TGRSzheng2020,IS_Ntubal}, such as the non-convex \cite{ChenGWWPH17,PengLKCWCKCC22,ChenHHZZZ22} and non-local \cite{ChenHYHZ20,XueZLC19,XueZLC19a} variants.

\subsection{Low-rank models combined with smoothness}
In addition to the LR prior, integrating local smoothness priors or regularizations \cite{LRTFDFR,LRTDTV} can greatly improve image reconstruction. Models such as TV-regularized SNN (SNNTV) \cite{SNNTV}, TV-regularized TNN (TNNTV) \cite{TNNTV}, and Smooth PARAFAC Completion with TV (SPCTV) \cite{SPCTV} are commonly used. However, these methods are sensitive to the TV weight, often requiring extensive parameter tuning \cite{E3DTV}. CTV and TCTV address this by modeling LR priors in the gradient domain, promoting both LR and smooth recovery in a regularization. However, TCTV is computationally intensive, taking over 50 minutes for a $512\times512\times63$ image. Furthermore, CTV and TCTV focus on high-frequency LR information, potentially overlooking important LR details in the image domain.

\section{Preliminaries}
\subsection{Tensor operations}
The Frobenius norm of a tensor $\tensor{A}\in\mathbb{R}^{I_1\times \cdots\times I_N}$ is defined as $\|\tensor{A}\|_{F}=\sqrt{\sum_{i_1,\cdots,i_N} |a_{i_1\cdots i_{N}}|^2}$. The mode-$n$ unfolding of a tensor results in a matrix representation, denoted by $\boldsymbol{A}_{(n)}\in\mathbb{R}^{I_{n}\times \prod_{i\neq n}I_{i}}$. The mode-$n$ product of a tensor $\tensor{A}\in\mathbb{R}^{I_1\times \cdots\times I_N}$ and a matrix $\boldsymbol{B}\in\mathbb{R}^{J\times I_n}$ results in a new tensor $\tensor{C}=\tensor{A}\times_{n}\boldsymbol{B}\in\mathbb{R}^{I_1\times \cdots\times I_{n-1}\times J\times I_{n+1}\times\cdots I_N}$, given by 
\begin{equation}
	c_{i_1,\cdots,i_{n-1},j,i_{n+1},\cdots,i_N} =\sum_{i_k=1}^{I_k} a_{i_1,\cdots,i_{n-1},i_n,i_{n+1},\cdots,i_N} b_{ji_n}.
\end{equation}
This can be represented in matrix form as $\boldsymbol{C}_{(n)} = \boldsymbol{B}\boldsymbol{A}_{(n)}.$

\begin{definition}[$n$-Rank \cite{n_Rank}] 
	The $n$-rank of a tensor $\tensor{A}$ is the rank of the tensor $\tensor{A}$ unfolded along mode $n$, given by $\mathrm{rank}_{n}(\tensor{A}) = \mathrm{rank}\left(\boldsymbol{A}_{(n)}\right)$.
\end{definition}

\begin{definition}[Tucker-Rank \cite{tucker_rank}]
	The Tucker-rank is a tuple of $n$-ranks, $\mathrm{rank}_{T}(\tensor{A}) =\left(\mathrm{rank}_{1}(\tensor{A}), \cdots, \mathrm{rank}_{N}(\tensor{A})\right)$.
\end{definition}

\subsection{Haar wavelet transform}
\begin{definition}[1-D Haar Discrete Wavelet Transform]
	The 1-D Haar discrete wavelet transform (HWT) of a vector $\boldsymbol{a}\in \mathbb{R}^N$ is defined by
	\begin{equation}
		\boldsymbol{b}=f(\boldsymbol{a})=\boldsymbol{W}_{N}\boldsymbol{a},
	\end{equation}
	where the $N$-order orthogonal projection matrix $\boldsymbol{W}_{N}$ is composed of an $\frac{N}{2}\times N$ block $\boldsymbol{H}_{N/2}$ and an $\frac{N}{2}\times N$ block $\boldsymbol{G}_{N/2}$, and the 1-D inverse HWT (IHWT) is given by
	\begin{equation}\label{eq:ihwt1}
		\boldsymbol{a} = f^{-1}(\boldsymbol{b})=\boldsymbol{W}_{N}^{-1}\boldsymbol{b}=\boldsymbol{W}_{N}^{T}\boldsymbol{b}.
	\end{equation}
\end{definition}
$\boldsymbol{H}_{N/2}$ and $\boldsymbol{G}_{N/2}$ are generally deemed as the average filter and discrete gradient operator, respectively; and $\boldsymbol{W}_{N}$ is defined by
\begin{equation}\label{eq:haar_wavelet_transform_matrix}
	\begin{aligned}
		\boldsymbol{W}_{N}  =& \left[ \begin{array}{c}
			\boldsymbol{H}_{N/2}\\
			\boldsymbol{G}_{N/2}\\
		\end{array} \right] \\
		=&\underbrace{\left[ \begin{matrix}
				\frac{\sqrt{2}}{2}&		\frac{\sqrt{2}}{2}&		0&		0&		\cdots&		0&		0\\
				0&		0&		\frac{\sqrt{2}}{2}&		\frac{\sqrt{2}}{2}&		\cdots&		0&		0\\
				\vdots&		&		&		&		\ddots&		&		\vdots\\
				0&		0&		0&		0&		\cdots&		\frac{\sqrt{2}}{2}&		\frac{\sqrt{2}}{2}\\  \hdashline
				\frac{\sqrt{2}}{2}&		-\frac{\sqrt{2}}{2}&		0&		0&		\cdots&		0&		0\\
				0&		0&		\frac{\sqrt{2}}{2}&		-\frac{\sqrt{2}}{2}&		\cdots&		0&		0\\
				\vdots&		&		&		&		\ddots&		&		\vdots\\
				0&		0&		0&		0&		\cdots&		\frac{\sqrt{2}}{2}&		-\frac{\sqrt{2}}{2}\\
			\end{matrix} \right]}_{N} ,
	\end{aligned}
\end{equation}

\begin{definition}[2-D Haar Discrete Wavelet Transform]
	The 2-D Haar discrete wavelet transform (2-D HWT) for a matrix $\boldsymbol{A}\in\mathbb{R}^{M\times N}$ is defined as
	\begin{equation}
		\boldsymbol{B}=F(\boldsymbol{A})=\boldsymbol{W}_{M}\boldsymbol{AW}_{N}^{T},
	\end{equation}
	where $\boldsymbol{W}_{M}$ and $\boldsymbol{W}_{N}$ are the $M$-order and $N$-order projection matrices, respectively. The 2-D inverse HWT (2-D IHWT) is given by
	\begin{equation}
		\boldsymbol{A}=F^{-1}(\boldsymbol{B})=\boldsymbol{W}_{M}^T\boldsymbol{B}\boldsymbol{W}_{N}.
	\end{equation}
\end{definition}

2-D HWT processes a matrix $\boldsymbol{A}\in\mathbb{R}^{M\times N}$ by applying the 1-D HWT to its row vectors and column vectors separately. Consequently, as depicted in Eq. (\ref{eq:haar_wavelet_transform_matrix}), $\boldsymbol{B}$ can be represented as a block matrix:
\begin{equation}
	\begin{aligned}
		\boldsymbol{B}=&\boldsymbol{W}_M\boldsymbol{A}\boldsymbol{W}_{N}^{T} \\
		=&\left[ \begin{array}{c}
			\boldsymbol{H}_{M/2}\\
			\boldsymbol{G}_{M/2}\\
		\end{array} \right] \boldsymbol{A}\left[ \begin{array}{c}
			\boldsymbol{H}_{N/2}\\
			\boldsymbol{G}_{N/2}\\
		\end{array} \right] ^T \\
		=&\left[ \begin{matrix}
			\boldsymbol{H}_{M/2}\boldsymbol{A}\boldsymbol{H}_{N/2}^{T}&		\boldsymbol{H}_{M/2}\boldsymbol{A}\boldsymbol{G}_{N/2}^{T}\\
			\boldsymbol{G}_{M/2}\boldsymbol{A}\boldsymbol{H}_{N/2}^{T}&		\boldsymbol{G}_{M/2}\boldsymbol{A}\boldsymbol{G}_{N/2}^{T}\\
		\end{matrix} \right] \\
		\triangleq&\left[ \begin{matrix}
			F_{1}(\boldsymbol{A})&		F_{2}(\boldsymbol{A})\\
			F_{3}(\boldsymbol{A})&		F_{4}(\boldsymbol{A})\\
		\end{matrix} \right] \triangleq\left[ \begin{matrix}
			\boldsymbol{B}_1&		\boldsymbol{B}_2\\
			\boldsymbol{B}_3&		\boldsymbol{B}_4\\
		\end{matrix} \right] ,
	\end{aligned}
\end{equation}
where $\boldsymbol{B}_1, \boldsymbol{B}_2, \boldsymbol{B}_3, \boldsymbol{B}_4$ represent the blocks known as approximation, horizontal, vertical, and diagonal wavelet coefficients, respectively.

The computational complexity of the 2-D HWT, as expressed by $\boldsymbol{B}=\boldsymbol{W}_M\boldsymbol{A}\boldsymbol{W}_{N}^T$, is $O(MN(M+N))$, which involves two instances of matrix multiplication. However, $\boldsymbol{W}_M$ and $\boldsymbol{W}_{N}$ are highly sparse matrices with distinctive patterns. The wavelet coefficients can be verified through the following expressions:
\begin{equation}\label{eq:wavelet_coefficient_index_sum}
	\begin{aligned}
		\boldsymbol{B}_1(i,j) &= \frac{a_{2i-1,2j-1}+a_{2i-1,2j}+a_{2i,2j-1}+a_{2i,2j}}{2}, \\
		\boldsymbol{B}_2(i,j) &= \frac{a_{2i-1,2j}+a_{2i,2j}-a_{2i-1,2j-1}-a_{2i,2j-1}}{2}, \\
		\boldsymbol{B}_3(i,j) &= \frac{a_{2i,2j-1}+a_{2i,2j}-a_{2i-1,2j-1}-a_{2i-1,2j}}{2}, \\
		\boldsymbol{B}_4(i,j) &= \frac{a_{2i-1,2j-1}+a_{2i,2j}-a_{2i-1,2j}+a_{2i,2j-1}}{2},
	\end{aligned}
\end{equation}
implicating that each wavelet coefficient is the sum of four $M/2 \times N/2$ matrices. Consequently, the computational complexity of the 2-D HWT via this method is diminished to $O(\frac34MN)$, with the 2-D IHWT exhibititing the same complexity of $O(\frac34MN)$ upon analogous considerations.

\begin{figure*}[t]
	\centering  
	\includegraphics[width=.9\linewidth]{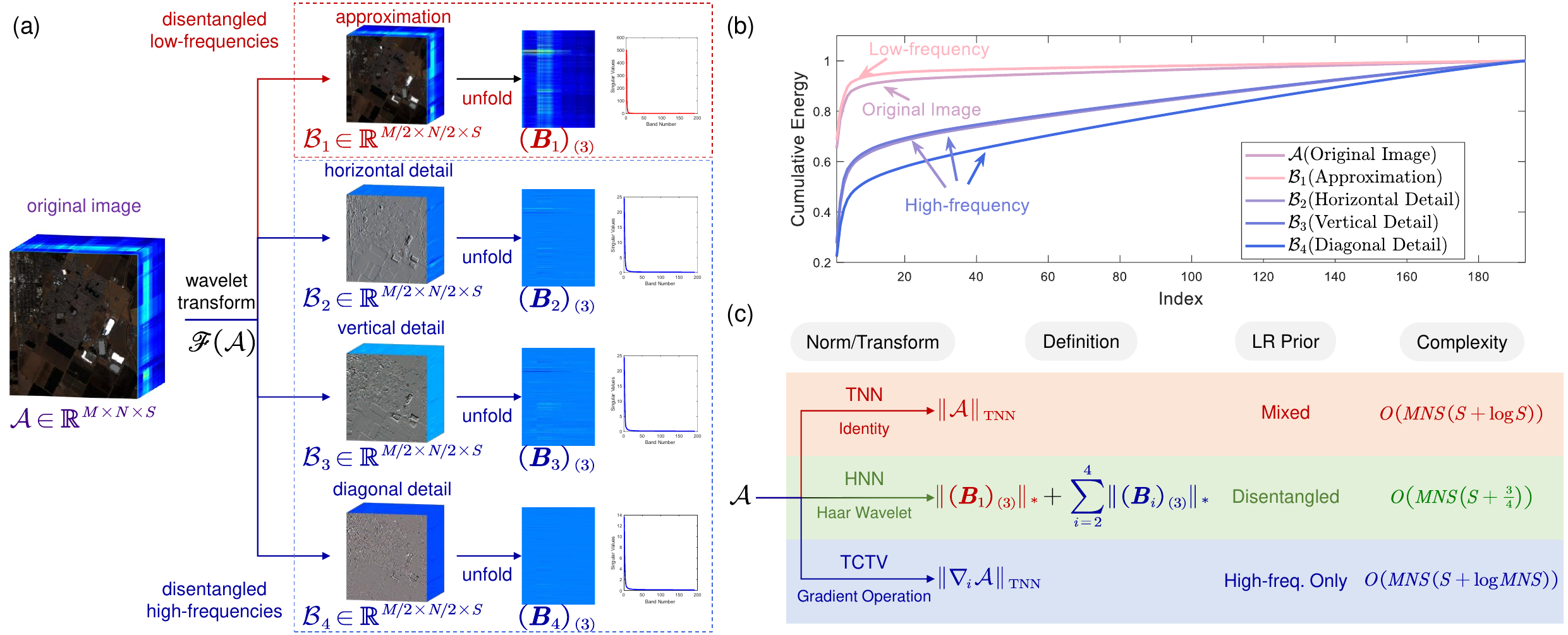}
	\caption{(a) Typical remote sensing imagery $\mathcal{A}\in \mathbb{R}^{M\times N\times S}$ and its wavelet coefficients, which exhibit rapid decay in singular values. $\mathscr{F}(\cdot)$ represents the 2D FHWT. (b) The CE curves for the original image and wavelet coefficients. (c) Comparison of TNN, TCTV, and HNN, where $\|\cdot\|_{*}$ denotes the matrix nuclear norm and $\|\cdot\|_{\rm TNN}$ represents the TNN. }
	\label{fig:wavelet_low_rank}
\end{figure*}

\section{Haar Nuclear Norm}
\subsection{Motivation}
The 2-D HWT decomposes an image into an approximation coefficient and three detail coefficients, and it has achieved significant advancements in image processing. However, it's unclear if these coefficients maintain low-rank properties for low-rank data, which this section explores.

The image considered in this paper is structured as a 3-order tensor, $\tensor{A} \in \mathbb{R}^{M \times N \times S}$, with $M \times N$ pixels and $S$ channels. Our focus is on characterizing spatial contents, which can be achieved through wavelet decomposition along the first and second modes. Consequently, the 2-D HWT for tensor images is defined as follows:
\begin{definition}[2-D Frontal Slice-wise Haar Discrete Wavelet Transform]
	Given a 3-order tensor $\tensor{A}\in\mathbb{R}^{M\times N\times S}$, the 2-D frontal slice-wise HWT (FHWT) is defined as:
	\begin{equation}
		\tensor{B}=\mathscr{F}(\tensor{A})=\tensor{A}\times_{1}\boldsymbol{W}_{M}\times_{2}\boldsymbol{W}_{N}.
	\end{equation}
	Equivalently, $\tensor{B}(:,:,k)=F\left(\tensor{A}(:,:,k)\right)=\boldsymbol{W}_{M}\tensor{A}(:,:,k)\boldsymbol{W}_{N}^{T}.$
	That is, 2-D FHWT is equivalent to applying 2-D HWT to each frontal slice of the tensor. The 2-D inverse FHWT (IFHWT) is given by: 
	\begin{equation}
		\tensor{A}=\mathscr{F}^{-1}(\tensor{B})=\tensor{B}\times_{1}\boldsymbol{W}_{M}^T\times_{2}\boldsymbol{W}_{N}^T.
	\end{equation}
\end{definition}
The wavelet coefficient $\tensor{B}$ can be represented as a block tensor comprising four wavelet coefficient tensors $\tensor{B}_{i} \in \mathbb{R}^{M/2 \times N/2 \times S}(i=1,2,3,4)$:
\begin{equation}\label{eq:wavelet_coefficient_block}
	\begin{aligned}
		\tensor{B}=&\tensor{A}\times_{1}\boldsymbol{W}_{M}\times_{2}\boldsymbol{W}_{N} \\
		=&\tensor{A}\times_{1} \left[ \begin{array}{c}
			\boldsymbol{H}_{M/2}\\
			\boldsymbol{G}_{M/2}\\
		\end{array} \right] \times_{2} \left[ \begin{array}{c}
			\boldsymbol{H}_{N/2}\\
			\boldsymbol{G}_{N/2}\\
		\end{array} \right] \\
		\triangleq&\left[ \begin{matrix}
			\mathscr{F}_{1}(\tensor{A})&		\mathscr{F}_{2}(\tensor{A})\\
			\mathscr{F}_{3}(\tensor{A})&		\mathscr{F}_{4}(\tensor{A})\\
		\end{matrix} \right]    \triangleq \left[ \begin{matrix}
			\tensor{B}_1&		\tensor{B}_2\\
			\tensor{B}_3&		\tensor{B}_4\\
		\end{matrix} \right] .
	\end{aligned}
\end{equation}
In this equation, $\mathscr{F}_{i}(\tensor{A})$ represents a mapping from tensor $\tensor{A}$ to the $i$-th wavelet coefficient tensor. It is noteworthy that the wavelet coefficients can be considered as the sum of four $M/2 \times N/2 \times S$ tensors (as detailed in the analysis near Eq. \ref{eq:wavelet_coefficient_index_sum}), thus the computational complexity of 2-D FHWT (or IFHWT) is $O(\frac34MNS)$.

If the original image is LR, its wavelet coefficients are intuitively also LR. This is supported by Fig. \ref{fig:wavelet_low_rank}(a), showing fast-decaying singular values across all wavelet coefficients. Furthermore, the following theorem delineates the low-rank property of the wavelet coefficients $\tensor{B}_{i}$:
\begin{theorem}[Low-rank Properties for Wavelet Coefficients]
	Consider a 3-order tensor $\tensor{A} \in \mathbb{R}^{M \times N \times S}$. Let $[\tensor{B}_{1}, \tensor{B}_{2}; \tensor{B}_{3}, \tensor{B}_{4}] = \mathrm{FHWT}_{2}(\tensor{A})$, and then there is 
	\begin{equation}
		\mathrm{rank}_{n}(\tensor{B}_{i}) \leq \mathrm{rank}_{n}(\tensor{A}),
	\end{equation}
	for $n = 1, 2, 3$ and $i = 1, 2, 3, 4$.
\end{theorem}
The proof is provided in the supplementary materials. The theorem states that the Tucker rank of the wavelet coefficients does not exceed that of the original data, confirming that wavelet coefficients remain LR if the original data is LR.

Cumulative energy (CE) curves provide more insights into the singular value distributions of wavelet coefficients. For a singular value sequence of length $n$, $(\lambda_{1},\ldots,\lambda_{n})$, the $k$-th CE is calculated as the ratio of the cumulative sum of the first $k$ singular values to the total sum of singular values, $\text{CE}_{k} = \frac{\sum_{i=1}^{k}\lambda_{i}}{\sum_{i=1}^{n}\lambda_{i}}$. Fig. \ref{fig:wavelet_low_rank}(b) shows that the CE curve for the approximation coefficient $\mathcal{B}_{1}$ rises faster than those for the detail coefficients $\mathcal{B}_{i}(i=2,3,4)$. This suggests that the singular values of low-frequencies are less dispersed, allowing for reconstruction with fewer principal components. In contrast, the singular values of high-frequencies are more dispersed, needing more components for reconstruction. Consequently, coarse-grained structures are more low-rank, while fine-grained details are less so. The original image, a mix of low- and high-frequencies, has a CE curve that falls between those of the low- and high-frequencies, indicating it is of intermediate low-rank.

NN/TNN applies LR regularization to the original image without considering the distinct singular value distributions of low- and high-frequencies, potentially leading to insufficient detail preservation. CTV/TCTV focuses on high-frequencies through gradient operators, ensuring their preservation but possibly inaccurately modeling the LR prior for coarse structures and risking color or spectral distortions.

\subsection{Haar nuclear norm}
This paper introduces the Haar Nuclear Norm (HNN) to model the disentangled LR properties of low- and high-frequency wavelet coefficients. 
\begin{definition}
	Given a 3-order tensor $\tensor{A}\in\mathbb{R}^{M\times N\times S}$, let $[\tensor{B}_{1},\tensor{B}_{2};\tensor{B}_{3},\tensor{B}_{4}]=\mathrm{FHWT}_{2}(\tensor{A})$, and HNN is defined as the sum of the nuclear norms of the mode-3 unfoldings of all wavelet coefficients,
	\begin{equation}
		\|\tensor{A}\|_{\mathrm{HNN}} = \sum_{i=1}^{4} \|\left(\boldsymbol{B}_{i}\right)_{(3)}\|_{*}.
	\end{equation}
\end{definition}

It is crucial to clarify the distinction between HNN and existing methods. Firstly, as illustrated in Fig. \ref{fig:wavelet_low_rank}, the existing tools primarily concentrate on LR recovery either in the original image domain (e.g., TNN) or the transformed domain (e.g., TCTV). The former is good at recovering coarse-grained structure but the entangled LR modeling may neglect detail preservation, while the latter maintains fine-grained details but may lack robust structure recovery. HNN differs from existing methods by explicitly modeling the LR properties of disentangled coarse-grained structures (approximation coefficients) and fine-grained textures (detail coefficients). It offers a balance between structure recovery and detail preservation, potentially outperforming methods like TNN and TCTV.

Despite involving the NN of four wavelet coefficients, HNN does not significantly increase computational burden. The most intensive operations are SVD and 2-D FHWT/IFHWT, with complexities of $O(MNS^2)$ and $O(\frac{3}{4}MNS)$, respectively. Since $\frac{3}{4}MNS \ll MNS^2$, the 2-D FHWT/IFHWT is negligible compared to SVD, making HNN only slightly slower than the NN in terms of computational complexity. However, TNN and TCTV involve additional complexities due to the fast Fourier transform, $O(MNS\log S)$ and $O(MNS\log MNS)$, respectively. Consequently, HNN is expected to be more computationally efficient than TNN and TCTV.

\subsection{HNN based matrix completion}

The HNN based matrix completion (HNN-MC) problem is formulated as 
\begin{equation}
	\min_{\tensor{X}} \|\tensor{X}\|_{\textrm{HNN}}, 
	\text { s.t. }  \mathscr{P}_{\Omega}(\tensor{M})=\mathscr{P}_{\Omega}(\tensor{X}).
\end{equation}
Note that remote sensing imagery is often written in a tensor format.
By introducing auxiliary variables $\tensor{E}$, it can be rewritten as 
\begin{equation}
	\left\{
	\begin{aligned}
		\min_{\tensor{X}}& \sum_{i=1}^{4} \|\left(\boldsymbol{B}_{i}\right)_{(3)}\|_{*}, \\
		\text { s.t. } & \tensor{M} = \tensor{X}+\tensor{E}, \mathscr{P}_{\Omega}(\tensor{E})=0, \tensor{B}_{i}=\mathscr{F}_{i}(\tensor{X}).
	\end{aligned}
	\right.
\end{equation}
Recall that $\mathscr{F}_{i}(\tensor{X})$ denotes a mapping from a tensor $\tensor{A}$ into the $i$-th wavelet coefficient, defined in Eq. (\ref{eq:wavelet_coefficient_block}).

The ADMM algorithm is then exploited to solve the issue, and the corresponding augmented Lagrangian function is 
\begin{equation}
	\begin{aligned}
		\mathscr{L} = & \left( \sum_{i=1}^{4}  \|\left(\boldsymbol{B}_{i}\right)_{(3)}\|_{*} + \frac{\mu_{b}}{2} \left\| \tensor{B}_{i}-\mathscr{F}_{i}(\tensor{X})+\frac{\Gamma_{i}}{\mu_{b}} \right\|_{F}^{2} \right) \\
		& + \frac{\mu_{a}}{2}\left\|\tensor{M} - \tensor{X}-\tensor{E}+\frac{\Gamma_{5}}{\mu_{a}} \right\|_{F}^{2},
	\end{aligned}
\end{equation}
with a constraint $\mathscr{P}_{\Omega}(\tensor{E})=0,$
where $\mu_{a}$ and $\mu_{b}$ are penalty parameters, and $\Gamma_{i} \in\mathbb{R}^{M/2\times N/2\times S} (i=1,2,3,4)$ and $\Gamma_{5} \in\mathbb{R}^{M\times N\times S}$ are Lagrangian multipliers. In the ADMM, each unknown variable is updated iteratively by fixing others, and the updating equations are discussed as follows:

Update $\tensor{E}$: Fixing other variables, the optimization problem of \mtensor{E} is given by 
\begin{equation}
	\min_{\tensor{E}}\left\|\tensor{E}-\left(\tensor{M}-\tensor{X}+\frac{\Gamma_5}{\mu_a}\right)\right\|_2^2, \text { s.t. } \mathscr{P}_{\Omega}(\tensor{E})=0,
\end{equation}
This is a constrained least squares problem, and the solution is given by
\begin{equation}\label{eq:update_E}
	\tensor{E} =  \mathscr{P}_{\Omega^{\perp}}\left(\tensor{M} - \tensor{X} + \frac{\Gamma_{5}}{\mu_{a}}\right),
\end{equation}
where $\Omega^{\perp}$ is the complement set of $\Omega$.

Update $\tensor{B}_{i}$: Fixing other variables, the optimization problem of $\tensor{B}_{i}$ is written as 
\begin{equation}
	\min_{\tensor{B}_{i}} 
	\|\left(\boldsymbol{B}_{i}\right)_{(3)}\|_{*} + \frac{\mu_{b}}{2} \left\| \tensor{B}_{i}-\mathscr{F}_{i}(\tensor{A}) +\frac{\Gamma_{i}}{\mu_{b}} \right\|_{F}^{2}. 
\end{equation}
The solution is given by singular value thresholding,
\begin{equation}\label{eq:update_B}
	\tensor{B}_i = \mathrm{fold}_{3}\left( 
	\boldsymbol{U}\mathscr{G}_{1/\mu_{b}}(\boldsymbol{\Sigma}) \boldsymbol{V}^T 
	\right),
\end{equation}
where $[\boldsymbol{U},\boldsymbol{\Sigma},\boldsymbol{V}^T]\triangleq\mathrm{svd}\left(\mathscr{F}_{i}(\tensor{A})-\Gamma_{i}/\mu_{b}\right)$, and $\mathscr{G}_{\gamma}(a) = \mathrm{sign}(a)\max(|a|-\gamma,0)$ denotes the soft-thresholding function.

Update \mtensor{X}: Fixing other variables, the optimization problem of \mtensor{X} is given by 
\begin{equation}
	\min_{\tensor{X}}   \frac{\mu_{b}}{2} \left\| \tensor{B}-\mathscr{F}(\tensor{X})+\frac{\Gamma}{\mu_{b}} \right\|_{F}^{2} + \frac{\mu_{a}}{2}\left\|\tensor{M} - \tensor{X}-\tensor{E}+\frac{\Gamma_{5}}{\mu_{a}} \right\|_{F}^{2}.
\end{equation}
This is a weighted least squares problem, and it is easy to compute the gradient of this loss function regarding $\tensor{X}$,
\begin{equation*}
	\begin{aligned}
		&\frac{\partial\mathscr{L}(\tensor{X})}{\partial\tensor{X}} \\
		=& \mu_a\left(\tensor{X}+\tensor{E}-\tensor{M}-\Gamma_5/\mu_a\right) +\mu_b(\tensor{X} \times_1 \boldsymbol{W}_M \times_2 \boldsymbol{W}_N\\
		& -\tensor{B}-\frac{\Gamma}{\mu_b}) \times_1 \boldsymbol{W}_M^{T} \times_2 \boldsymbol{W}_N^{T} \\
		=& \mu_a\left(\tensor{X}+\tensor{E}-\tensor{M}-\Gamma_5/\mu_a\right) +\mu_b\tensor{X} - \mathscr{F}^{-1}\left(\mu_{b}\tensor{B}+\Gamma\right) .
	\end{aligned} 
\end{equation*}
Let the gradient be zero, and the solution is 
\begin{equation}
	\tensor{X} = \frac{\mu_{a}(\tensor{M}+\Gamma_{5}/\mu_{a}-\tensor{E})+\mathscr{F}^{-1}\left(\mu_b\tensor{B}+\Gamma\right) }{\mu_{a}+\mu_{b}}.
\end{equation}

Except for the aforementioned variables, the multipliers and penalty parameters are updated by 
\begin{equation}
	\begin{aligned}
		\Gamma_i&=\Gamma _i+\mu _b\left( \tensor{B}_{i}-\mathscr{F}_i\left( \tensor{X} \right)  \right), \,\, \left( i=1,2,...,4 \right) \\
		\Gamma_5&=\Gamma _5+\mu _a\left( \tensor{M}-\tensor{X}-\tensor{E} \right), \, \mu_a=\mu _a\rho \,\,,\,\, \mu _b=\mu _b\rho ,
	\end{aligned}
\end{equation}
where $\rho$ is constant greater than 1.

\subsection{HNN based robust principal component analysis}
Given a noisy observation $\tensor{M}$, it is assumed that $\tensor{M}$ can be decomposed into a low-rank component $\tensor{X}$ (i.e., clean data) and a sparse noise component $\tensor{S}$. The HNN based robust principal component analysis (HNN-RPCA) problem tries to recover $\tensor{X}$, formulated as 
\begin{equation}
	\min_{\tensor{X}} \|\tensor{X}\|_{\textrm{HNN}} + \lambda \|\tensor{E}\|_{1},
	\text { s.t. }  \tensor{M}=\tensor{X}+\tensor{E}.
\end{equation}
With the definition of HNN, it is rewritten as 
\begin{equation}
	\left\{
	\begin{aligned}
		\min_{\tensor{X}}& \sum_{i=1}^{4} \|\left(\boldsymbol{B}_{i}\right)_{(3)}\|_{*}+ \lambda \|\tensor{E}\|_{1}, \\
		\text { s.t. } & \tensor{M} = \tensor{X}+\tensor{E},  \tensor{B}_{i}=\mathscr{F}_{i}(\tensor{X}).
	\end{aligned}
	\right.
\end{equation}
Its augmented Lagrangian function is 
\begin{equation}
	\begin{aligned}
		\mathscr{L} = & \left( \sum_{i=1}^{4}  \|\left(\boldsymbol{B}_{i}\right)_{(3)}\|_{*} + \frac{\mu_{b}}{2} \left\| \tensor{B}_{i}-\mathscr{F}_{i}(\tensor{X})+\frac{\Gamma_{i}}{\mu_{b}} \right\|_{F}^{2} \right) \\
		&+ \lambda \|\tensor{E}\|_{1} + \frac{\mu_{a}}{2}\left\|\tensor{M} - \tensor{X}-\tensor{E}+\frac{\Gamma_{5}}{\mu_{a}} \right\|_{F}^{2},
	\end{aligned}
\end{equation}
where $\mu_{a}$ and $\mu_{b}$ are penalty parameters, and $\Gamma_{i} \in\mathbb{R}^{M/2\times N/2\times S} (i=1,2,3,4)$ and $\Gamma_{5} \in\mathbb{R}^{M\times N\times S}$ are Lagrangian multipliers. The solution is very similar to that of HNN based matrix completion, as follows:
\begin{equation}
	\left\{
	\begin{aligned}
		\tensor{E} & =  \mathscr{G}_{\lambda/\mu_{a}}\left(\tensor{M} - \tensor{X} + \frac{\Gamma_{5}}{\mu_{a}}\right), \\
		\tensor{B}_i & = \mathrm{fold}_{3}\left( 
		\boldsymbol{U}\mathscr{G}_{1/\mu_{b}}(\boldsymbol{\Sigma}) \boldsymbol{V}^T 
		\right), \\
		\tensor{X} & = \frac{\mu_{a}(\tensor{M}+\Gamma_{5}/\mu_{a}-\tensor{E})+\mathscr{F}^{-1}\left(\mu_b\tensor{B}+\Gamma\right) }{\mu_{a}+\mu_{b}}, \\
		\Gamma_i&=\Gamma _i+\mu _b\left( \tensor{B}_{i}-\mathscr{F}_i\left( \tensor{X} \right)  \right), \,\, \left( i=1,2,...,4 \right) \\
		\Gamma_5&=\Gamma _5+\mu _a\left( \tensor{M}-\tensor{X}-\tensor{E} \right), \\
		\mu_a&=\mu _a\rho \,\,,\,\, \mu _b=\mu _b\rho .
	\end{aligned}
	\right.
\end{equation}

\section{Theory of Recovery Guarantee}
The section develops the theory of recovery guarantee for HNN-RPCA and HNN-TC models. To ease representation, this section would use the matrix notations. Firstly, incoherence conditions on the transformed maps  after Haar transformation are defined. Specifically, assuming that each transformed map $(\boldsymbol{B}_{i})_{(3)}\in\mathbb{R}^{MN/4\times S}, (i=1,\cdots,4)$ of $\bm{X}_0\in\mathbb{R}^{MN\times S}$ with rank of $r$ has the singular value decomposition $\bm{U}_i\bm{\Sigma}_i\bm{V}_i^T$, where $\bm{U}_i\in \mathbb{R}^{MN\times r}$ and $\bm{V}_i\in \mathbb{R}^{S\times r}$, $n_1=MN/4$ and $n_2=S$, the incoherence condition with a fixed constant $\mu$ is defined as follows:
\begin{equation}
	\label{in_U}
	\max _k\left\|\bm{U}_i^T \hat{e}_k\right\|^2 \leq \frac{\mu r}{n_1}, i=1,\cdots,4. 
\end{equation}
\begin{equation}
	\label{in_V}
	\max _k\left\|\bm{V}_i^T\hat{e}_k\right\|^2\leq \frac{\mu r}{n_2}, i=1,\cdots,4.
\end{equation}
\begin{equation}
	\label{in_UV}
	\|\bm{U}_i \bm{V}_i^T\|_{\infty} \leq \sqrt{\frac{\mu r}{n_1 n_2}}, i=1,\cdots,4 .
\end{equation}
Here, $\Vert \bm{X}\Vert_{\infty} = \max \limits_{i,j} |\bm{X}_{i,j}|$, and $\hat{e}_k$ is the standard orthonormal basis. The exact decomposition theorem for HNN-RPCA model is derived as follows, where we now write it in matrix notations, 
\begin{equation}\label{rpca_hnn}
	\min_{\bm{X}}  \|\bm{X}\|_{\textrm{HNN}} + \lambda \|\bm{E}\|_{1},   \textrm{ s.t. }  \bm{M}=\bm{X}+\bm{E}.
\end{equation}

\begin{figure*}[]
	\centering
	\subfigure[RPCA]{\includegraphics[width=0.4\linewidth]{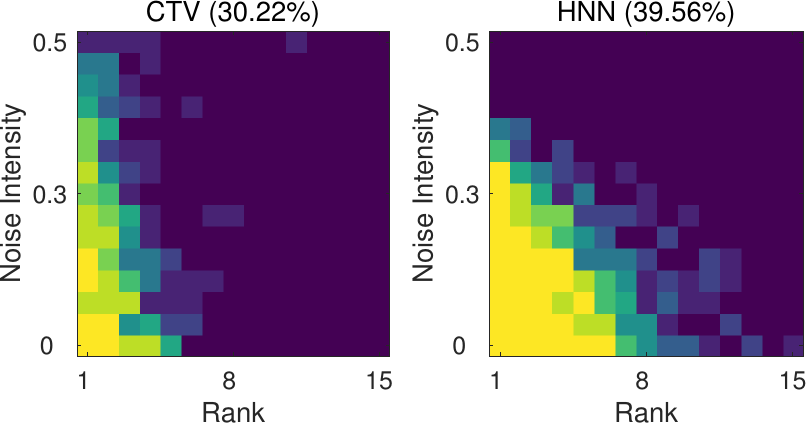}}
	\subfigure[MC]{\includegraphics[width=0.4\linewidth]{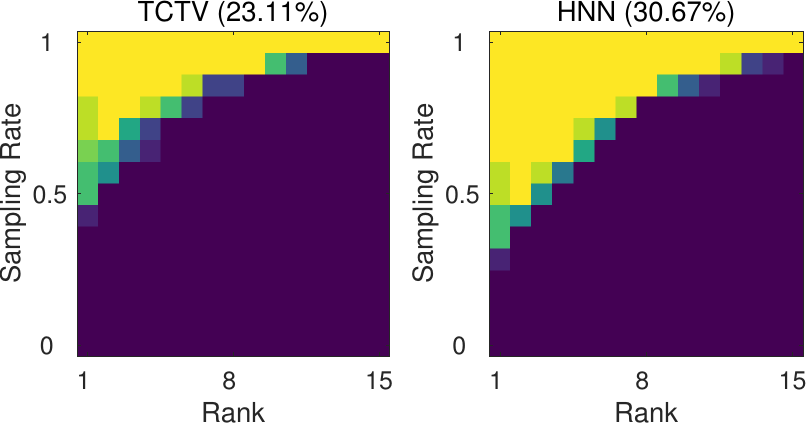}}
	\caption{Phase transitions of (a) RPCA models and (b) MC models. The success rates are shown in the title of each panel. }
	\label{fig:phase_map}
\end{figure*}
\begin{figure*}[]
	\centering
	\includegraphics[width=.85\linewidth]{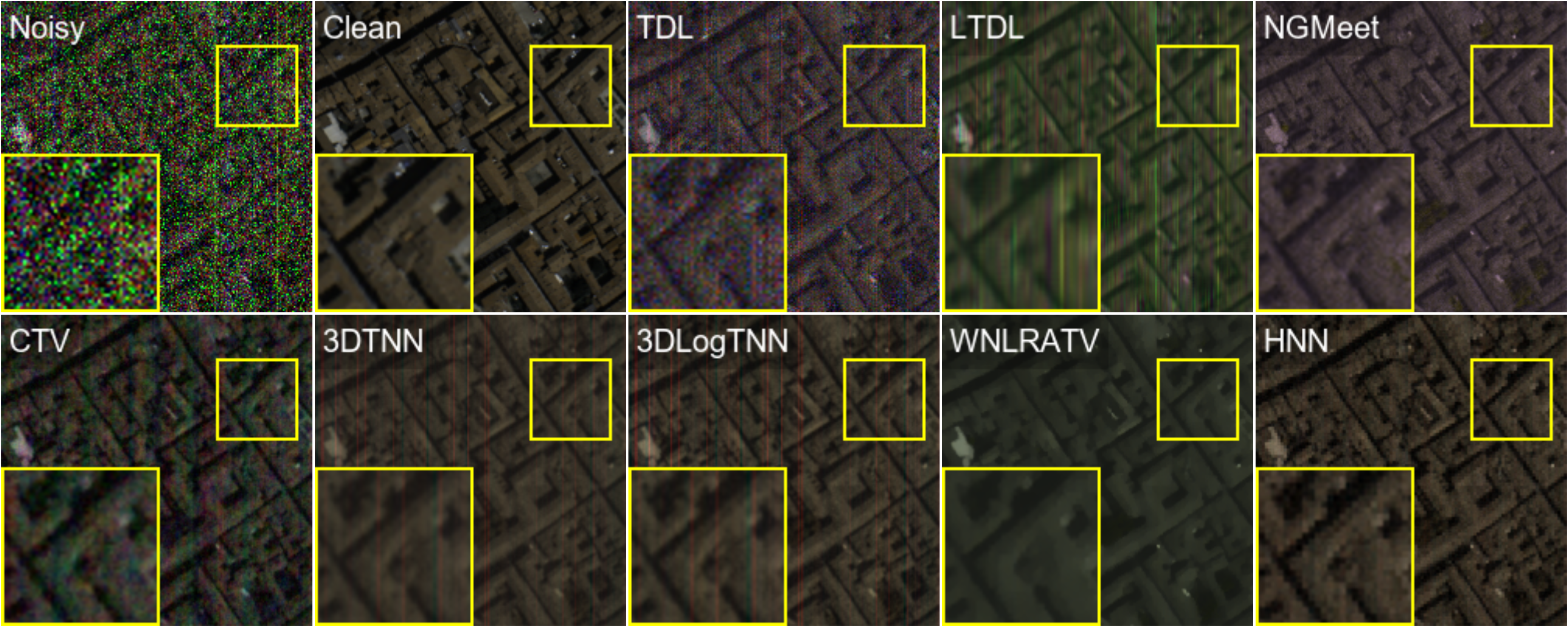}
	\caption{The false-color denoised images (band 24-17-1) of compared methods on the PC dataset for case 6.}
	\label{fig:pc_case6}
\end{figure*}

\begin{table*}[tbp]
	\centering
	\caption{HSI denoising metrics on the BA dataset. The best and the second best results are highlighted in \textbf{bold} and {\ul underline}, respectively. }
	\resizebox{0.9\textwidth}{!}{
\begin{tabular}{ccccccccccccc}
	\hline
	Case  & \multicolumn{1}{c}{Metrics} & \multicolumn{1}{c}{Noisy} & \multicolumn{1}{c}{LLRT} & \multicolumn{1}{c}{TDL} & \multicolumn{1}{c}{LTDL} & \multicolumn{1}{c}{NGMeet} & \multicolumn{1}{c}{CTV} & \multicolumn{1}{c}{3DTNN} & \multicolumn{1}{c}{3DLogTNN} & \multicolumn{1}{c}{WNLRATV} & \multicolumn{1}{c}{RCILD} & \multicolumn{1}{c}{HNN} \bigstrut[b]\\
	\hline
	\multirow{4}[2]{*}{1} & PSNR$\uparrow$ & 13.06 & 15.6  & 22.9  & 23.04 & 23.15 & 27.95 & 26.7  & 24.42 & \underline{28.37} & 23.22 & \textbf{31.89} \bigstrut[t]\\
	& SSIM$\uparrow$ & 0.0308 & 0.062 & 0.6801 & 0.7085 & 0.7408 & 0.4991 & 0.5886 & 0.3614 & \underline{0.7549} & 0.7381 & \textbf{0.7864} \\
	& ERGAS$\downarrow$ & 1114.99 & 838.17 & 416.67 & 414.47 & 411.94 & \underline{196.53} & 256.8 & 303.19 & 210.78 & 411.8 & \textbf{126.5} \\
	& SAM$\downarrow$  & 0.269 & 0.217 & 0.083 & 0.082 & 0.081 & 0.091 & \underline{0.080} & 0.108 & 0.087 & 0.082 & \textbf{0.045} \bigstrut[b]\\
	\hline
	\multirow{4}[2]{*}{2} & PSNR$\uparrow$ & 14.3  & 16.63 & 24.18 & 25.66 & 24.38 & \underline{29.17} & 28.15 & 26.36 & 28.84 & 26.13 & \textbf{32.16} \bigstrut[t]\\
	& SSIM$\uparrow$ & 0.0467 & 0.0847 & 0.5333 & 0.748 & 0.7006 & 0.5677 & 0.6416 & 0.4666 & 0.7405 & \underline{0.7892} & \textbf{0.8064} \\
	& ERGAS$\downarrow$ & 1046.25 & 818.79 & 412.75 & 365.74 & 406.52 & \underline{174.26} & 230.81 & 277.38 & 219.36 & 361.53 & \textbf{129.73} \\
	& SAM$\downarrow$  & 0.266 & 0.221 & 0.099 & 0.086 & 0.100 & 0.080 & \underline{0.076} & 0.106 & 0.092 & 0.085 & \textbf{0.047} \bigstrut[b]\\
	\hline
	\multirow{4}[2]{*}{3} & PSNR$\uparrow$ & 13.33 & 15.77 & 22.82 & 23.77 & 23.12 & \underline{29.23} & 25.05 & 21.35 & 28.48 & 24.91 & \textbf{32.02} \bigstrut[t]\\
	& SSIM$\uparrow$ & 0.0408 & 0.0772 & 0.4622 & 0.7111 & 0.6552 & 0.5706 & 0.329 & 0.1923 & 0.7041 & \underline{0.7612} & \textbf{0.802} \\
	& ERGAS$\downarrow$ & 1195.65 & 932.48 & 471.94 & 436.95 & 463.06 & \underline{170.52} & 295.23 & 476.74 & 248.07 & 409.96 & \textbf{130.87} \\
	& SAM$\downarrow$  & 0.278 & 0.232 & 0.108 & 0.096 & 0.107 & \underline{0.081} & 0.119 & 0.178 & 0.106 & 0.094 & \textbf{0.049} \bigstrut[b]\\
	\hline
	\multirow{4}[2]{*}{4} & PSNR$\uparrow$ & 14.35 & 16.68 & 23.96 & 25.36 & 24.6  & 29.1  & 27.45 & 24.89 & \underline{29.58} & 26.52 & \textbf{32.26} \bigstrut[t]\\
	& SSIM$\uparrow$ & 0.0473 & 0.0862 & 0.5004 & 0.742 & 0.7169 & 0.564 & 0.5438 & 0.3648 & 0.7839 & \underline{0.7972} & \textbf{0.8079} \\
	& ERGAS$\downarrow$ & 1024.37 & 797.55 & 401.08 & 362.24 & 386.97 & \underline{174.07} & 234.16 & 313.58 & 198.36 & 336.48 & \textbf{125.28} \\
	& SAM$\downarrow$  & 0.265 & 0.220 & 0.099 & 0.084 & 0.096 & \underline{0.080} & 0.084 & 0.122 & 0.084 & 0.082 & \textbf{0.046} \bigstrut[b]\\
	\hline
	\multirow{4}[2]{*}{5} & PSNR$\uparrow$ & 14.86 & 17.22 & 21.92 & 23.48 & 25.74 & 27.35 & 26.74 & \underline{27.42} & 26.34 & 26.72 & \textbf{29.16} \bigstrut[t]\\
	& SSIM$\uparrow$ & 0.0525 & 0.0971 & 0.2834 & 0.6506 & \underline{0.7111} & 0.5363 & 0.6656 & 0.6634 & 0.567 & \textbf{0.7472} & 0.6868 \\
	& ERGAS$\downarrow$ & 978.93 & 766.22 & 472.17 & 273.05 & 353.64 & \underline{235.23} & 267.21 & 256.9 & 314.17 & 309.07 & \textbf{219.12} \\
	& SAM$\downarrow$  & 0.276 & 0.233 & 0.154 & \textbf{0.084} & 0.111 & 0.123 & 0.108 & 0.110 & 0.148 & \underline{0.095} & 0.105 \bigstrut[b]\\
	\hline
	\multirow{4}[2]{*}{6} & PSNR$\uparrow$ & 13.47 & 15.97 & 20.75 & 21.46 & 24.1  & \underline{27.25} & 26.45 & 25.35 & 25.78 & 25.11 & \textbf{28.99} \bigstrut[t]\\
	& SSIM$\uparrow$ & 0.0417 & 0.0817 & 0.2425 & 0.5546 & 0.6633 & 0.5288 & 0.5819 & 0.4671 & 0.5201 & \textbf{0.7145} & \underline{0.67} \\
	& ERGAS$\downarrow$ & 1154.8 & 892.02 & 524.67 & 332.13 & 422.28 & \underline{234.78} & 260.3 & 298.07 & 368.66 & 380.11 & \textbf{224.18} \\
	& SAM$\downarrow$  & 0.288 & 0.242 & 0.153 & \textbf{0.096} & 0.115 & 0.122 & 0.110 & 0.131 & 0.171 & 0.111 & \underline{0.108} \bigstrut[b]\\
	\hline
\end{tabular}%

}
	\label{tab:BA_denoisnig}%
\end{table*}%

\begin{theorem}[HNN-RPCA Theorem]
	\label{main_theorem_rpca}
	Suppose that the transformed map $(\boldsymbol{B}_{i})_{(3)}\in\mathbb{R}^{MN/4\times S}, (i=1,\cdots,4)$ of $\bm{X}_0\in\mathbb{R}^{MN\times S}$ with rank $r$ satisfies the incoherence conditions (\ref{in_U})-(\ref{in_UV}), and the support set $\Omega$ of $\bm{E}_0$ is uniformly distributed over all sets of cardinality $m$. Then, with a probability of at least $1-cn_{(1)}^{-10}$ (for the choice of support set exceeding $\bm{E}_0$), the HNN-RPCA model with $\lambda = 4/\sqrt{n_{(1)}}$ guarantees an exact recovery, i.e., the minimization in Eq. (\ref{rpca_hnn}) yields $\hat{\bm{X}} = \bm{X}_0$ and $\hat{\bm{E}} = \bm{E}_0$, provided that:
	\begin{equation}
		\label{rpca_cond}
		\operatorname{rank}\left(\bm{X}_0\right) \leq \rho_r n_2 \mu^{-1}\left(\log n_{(1)}\right)^{-2} ,m \leq \rho_s n_1 n_2.
	\end{equation}
	Here, $\rho_r$ and $\rho_s$ are numerical constants, where $\rho_s$ determines the sparsity of $\bm{E}_0$, $\rho_r$ is a small constant related to the rank of $\bm{X}_0$, and $n_{(1)}=\max \{ MN/4, S\}$.
\end{theorem}
Similarly, for the HNN-MC model which is written in matrix notations as: 
\begin{equation}
	\label{hnn_mc}
	\min_{\bm{X}} \ \|\bm{X}\|_{\textrm{HNN}}, \text { s.t. } \mathscr{P}_{\Omega}(\bm{M})=\mathscr{P}_{\Omega}(\bm{X}),
\end{equation}
there is the following theorem. 
\begin{theorem}[HNN-MC Theorem]
	\label{theorem_main4}
	Suppose that $\bm{B}_i \in \mathbb{R}^{MN/4\times S}, (i=1,\cdots,4)$ obey the incoherence conditions (\ref{in_U})-(\ref{in_V}), $\Omega\sim\mbox{Ber}(p)$ and $ m$ is the number of $\Omega $, where $\mbox{Ber}(p)$ represents the Bernoulli distribution with parameter $p$. Let $n_1=MN/4$ and $n_2=S$, and without loss of generality, suppose $ n_1 \geq n_2 $. Then, there exist universal constants $c_0, c_1>0$ such that $\bm{X}_0$ is the unique solution to MNN-MC model (\ref{hnn_mc}) with probability at least $1-c_1n_1^{-3}\log n_1$, provided that
	\begin{equation}\label{eq.17_mnn}
		m\geq c_0\mu r n_1^{5/4}\log(n_1).
	\end{equation}
\end{theorem}
The proofs of the above theorems are placed in the supplementary material. These theorems show that the HNN-RPCA/-MC model has the exact recovery with a high probability.

\section{Experiments}

The proposed model's effectiveness is tested through extensive experiments using four quality metrics: Peak Signal-to-Noise Ratio (PSNR), Structural Similarity (SSIM), Erreur Relative Globale Adimensionnelle de Synthese (ERGAS), and Spectral Angle Mapper (SAM). PSNR, SSIM, and ERGAS evaluate spatial distortion, while SAM measures spectral distortion. Better image quality is indicated by higher PSNR and SSIM scores, and lower ERGAS and SAM scores. Experiments were conducted on a Windows 11 desktop with an 8-core R7-4800H CPU and an RTX 3060 GPU (12 GB GPU memory).

\begin{table*}[tbp]
	\centering
	\caption{HSI denoising metrics on the PC dataset. The best and the second best results are highlighted in \textbf{bold} and {\ul underline}, respectively.}
	\resizebox{0.9\textwidth}{!}{
\begin{tabular}{ccccccccccccc}
	\hline
	Case  & \multicolumn{1}{c}{Metrics} & \multicolumn{1}{c}{Noisy} & \multicolumn{1}{c}{LLRT} & \multicolumn{1}{c}{TDL} & \multicolumn{1}{c}{LTDL} & \multicolumn{1}{c}{NGMeet} & \multicolumn{1}{c}{CTV} & \multicolumn{1}{c}{3DTNN} & \multicolumn{1}{c}{3DLogTNN} & \multicolumn{1}{c}{WNLRATV} & \multicolumn{1}{c}{RCILD} & \multicolumn{1}{c}{HNN} \bigstrut[b]\\
	\hline
	\multirow{4}[2]{*}{1} & PSNR$\uparrow$ & 12.54 & 16.96 & 23.35 & 23.04 & 23.92 & 25.95 & 25.12 & \underline{27.04} & 24.02 & 24.04 & \textbf{27.59} \bigstrut[t]\\
	& SSIM$\uparrow$ & 0.117 & 0.255 & 0.7348 & 0.7085 & 0.7947 & 0.7116 & 0.7288 & \underline{0.8074} & 0.7155 & 0.8061 & \textbf{0.8099} \\
	& ERGAS$\downarrow$ & 865.04 & 521.08 & 255.24 & 414.47 & 241.74 & 185.14 & 204.08 & \underline{165.22} & 233.17 & 238.13 & \textbf{152.69} \\
	& SAM$\downarrow$  & 0.243 & 0.155 & 0.047 & 0.082 & 0.043 & 0.081 & \underline{0.042} & \textbf{0.039} & 0.047 & 0.043 & 0.044 \bigstrut[b]\\
	\hline
	\multirow{4}[2]{*}{2} & PSNR$\uparrow$ & 14.17 & 17.98 & 22.66 & 25.66 & 24.37 & 27.6  & 27.13 & \underline{28.16} & 27.45 & 25.94 & \textbf{28.45} \bigstrut[t]\\
	& SSIM$\uparrow$ & 0.1789 & 0.3117 & 0.5653 & 0.748 & 0.7478 & 0.783 & 0.7706 & 0.7953 & 0.8035 & \textbf{0.8471} & \underline{0.8452} \\
	& ERGAS$\downarrow$ & 773.26 & 505.14 & 302.96 & 365.74 & 257.31 & \underline{155.13} & 166.15 & 157.8 & 178.52 & 207.37 & \textbf{139.07} \\
	& SAM$\downarrow$  & 0.234 & 0.163 & 0.091 & 0.086 & 0.073 & 0.070 & 0.054 & 0.062 & 0.056 & \underline{0.049} & \textbf{0.044} \bigstrut[b]\\
	\hline
	\multirow{4}[2]{*}{3} & PSNR$\uparrow$ & 13.5  & 17.39 & 22.05 & 23.77 & 23.44 & 27.86 & 27.31 & \textbf{28.36} & 27.06 & 24.86 & \underline{28.24} \bigstrut[t]\\
	& SSIM$\uparrow$ & 0.1609 & 0.2901 & 0.535 & 0.7111 & 0.7168 & 0.794 & 0.7805 & 0.8013 & 0.7725 & \underline{0.8227} & \textbf{0.8404} \\
	& ERGAS$\downarrow$ & 842.82 & 554.27 & 326.73 & 436.95 & 316.19 & \underline{150.55} & 162.93 & 161.15 & 254.39 & 252.96 & \textbf{145.68} \\
	& SAM$\downarrow$  & 0.242 & 0.173 & 0.093 & 0.096 & 0.086 & 0.069 & \underline{0.054} & 0.065 & 0.079 & 0.061 & \textbf{0.048} \bigstrut[b]\\
	\hline
	\multirow{4}[2]{*}{4} & PSNR$\uparrow$ & 13.12 & 17.28 & 22.08 & 25.36 & 23.7  & 26.45 & 25.17 & \underline{27.03} & 25.25 & 25.73 & \textbf{27.76} \bigstrut[t]\\
	& SSIM$\uparrow$ & 0.1369 & 0.2721 & 0.5492 & 0.742 & 0.7483 & 0.7374 & 0.7312 & 0.8083 & 0.7293 & \textbf{0.8391} & \underline{0.8186} \\
	& ERGAS$\downarrow$ & 835.9 & 521.79 & 304.17 & 362.24 & 259.7 & 175.11 & 203.36 & \underline{165.42} & 214.67 & 214.09 & \textbf{149.84} \\
	& SAM$\downarrow$  & 0.241 & 0.161 & 0.081 & 0.084 & 0.064 & 0.078 & \underline{0.044} & \textbf{0.042} & 0.056 & 0.052 & 0.045 \bigstrut[b]\\
	\hline
	\multirow{4}[2]{*}{5} & PSNR$\uparrow$ & 13.52 & 17.6  & 19.86 & 23.55 & 24.02 & 24.13 & 23.91 & 25.11 & 25.16 & \textbf{25.6} & \underline{25.43} \bigstrut[t]\\
	& SSIM$\uparrow$ & 0.1468 & 0.28  & 0.3895 & 0.5351 & 0.7315 & 0.662 & 0.6469 & 0.6957 & 0.7094 & \textbf{0.7664} & \underline{0.7447} \\
	& ERGAS$\downarrow$ & 813.08 & 510.51 & 399.92 & 406.29 & 255.25 & 263.17 & 266.18 & 252.27 & 242.79 & \textbf{227.82} & \underline{241.2} \\
	& SAM$\downarrow$  & 0.252 & 0.173 & 0.136 & 0.124 & \underline{0.080} & 0.115 & 0.092 & 0.096 & 0.087 & \textbf{0.080} & 0.094 \bigstrut[b]\\
	\hline
	\multirow{4}[2]{*}{6} & PSNR$\uparrow$ & 12.96 & 17.12 & 19.21 & 22.3  & 23.56 & 24.02 & 23.29 & \underline{24.23} & 23.64 & 23.79 & \textbf{24.88} \bigstrut[t]\\
	& SSIM$\uparrow$ & 0.1353 & 0.2645 & 0.3528 & 0.472 & \underline{0.7248} & 0.6633 & 0.6065 & 0.6518 & 0.5969 & 0.7171 & \textbf{0.7288} \\
	& ERGAS$\downarrow$ & 875.19 & 552.58 & 433.47 & 460.26 & 287.06 & 270.3 & 280.32 & 271.1 & 287.3 & \textbf{259.8} & \underline{260.4} \\
	& SAM$\downarrow$  & 0.260 & 0.183 & 0.145 & 0.133 & \underline{0.092} & 0.119 & 0.098 & 0.105 & 0.092 & \textbf{0.085} & 0.108 \bigstrut[b]\\
	\hline
\end{tabular}%
}
	\label{tab:PC_denoising}%
\end{table*}%

\subsection{Experiments on phase transition}
The recovery phenomenon was analyzed across varying Tucker ranks and noise intensities of $\bm{E}$ for RPCA problems, or varying sampling rates of $\bm{M}$ for MC problems. A simulated tensor $\tensor{X} \in \mathbb{R}^{30 \times 30 \times 30}$ was generated using the equation $\tensor{X} = \tensor{C} \times_1 \bm{U}_{1} \times_2 \bm{U}_{2} \times_3 \bm{U}_{3}$, where $\tensor{C} \in \mathbb{R}^{R \times R \times R}$ and $\bm{U}_{i} \in \mathbb{R}^{30 \times R}$, for $i = 1, 2, 3$. The rank $R$ varied from 1 to 30, the noise intensity ranged from 0.01 to 0.5, and the sampling rate varied from 0.01 to 0.99. For each case, the experiment was repeated 10 times, and a trial was deemed successful if the average relative reconstruction error was less than 0.1. Figure \ref{fig:phase_map} illustrates the phase transition maps (yellow = 100\%, blue = 0\%), indicating that HNN achieved a higher success rate than CTV and TCTV.

\begin{table*}[!t]
	\renewcommand{\arraystretch}{1.3}
	\caption{Metrics for HSI inpainting on the BA dataset. The best and the second best results are highlighted in \textbf{bold} and {\ul underline}, respectively.}
	\label{metrics for the BayArea HSIs dataset}
	\centering
	\resizebox{0.9\textwidth}{!}{
	\begin{tabular}{ccccccccccccc}
		\hline
		
		SR                   & Metrics & Observed & LRTFR & SNN            & KBR          & TNN    & SPCTV   & TNNTV   & FCTNTC & FCTNFR & TCTV         & HNN             \\ \hline
		\multirow{4}{*}{7\%} & PSNR$\uparrow$   & 20.09    & 39.07  & 31.82          & {\ul 44.88}  & 39.47  & 40.07   & 34.05   & 37.06  & 39.89  & 44.43        & \textbf{48.46}  \\
		& SSIM$\uparrow$   & 0.0864   & 0.9216 & 0.8254         & 0.9742       & 0.9314 & 0.9492  & 0.8542   & 0.8920 & 0.9385 & {\ul 0.9749} & \textbf{0.9898} \\
		& ERGAS$\downarrow$  & 535.81   & 55.11  & 138.69         & {\ul 31.90}  & 61.77  & 52.10   & 108.97  & 73.48  & 52.66  & 33.26        & \textbf{22.30}  \\
		& SAM$\downarrow$   & 1.3075   & 0.0595 & 0.1374         & {\ul 0.0397} & 0.0857 & 0.0632  & 0.0962   & 0.0859 & 0.0643 & 0.0456       & \textbf{0.0288} \\ \hline
		\multirow{4}{*}{5\%} & PSNR$\uparrow$   & 20.00    & 38.88  & 31.10          & 41.46        & 36.83  & 37.67   & 33.15   & 36.89  & 39.40  & {\ul 42.40}  & \textbf{45.57}  \\
		& SSIM$\uparrow$   & 0.0708   & 0.9181 & 0.8145         & 0.9501       & 0.8908 & 0.9215  & 0.8412   & 0.8883 & 0.9316 & {\ul 0.9616} & \textbf{0.9824} \\
		& ERGAS$\downarrow$  & 541.57   & 56.61  & 150.60         & 47.38        & 85.19  & 69.13   & 121.79  & 74.92  & 55.64  & {\ul 43.01}  & \textbf{30.77}  \\
		& SAM$\downarrow$   & 1.3515   & 0.0637 & 0.1489         & 0.0534       & 0.1173 & 0.0802  & 0.1059   & 0.0871 & 0.0669 & {\ul 0.0576} & \textbf{0.0396} \\ \hline
		\multirow{4}{*}{4\%} & PSNR$\uparrow$   & 19.95    & 38.63  & 30.74          & 39.20        & 35.17  & 36.14   & 32.70   & 36.72  & 39.05  & {\ul 39.76}  & \textbf{43.00}  \\
		& SSIM$\uparrow$   & 0.0623   & 0.9139 & 0.8097         & 0.9267       & 0.8587 & 0.8993  & 0.8343   & 0.8852 & 0.9258 & {\ul 0.9375} & \textbf{0.9688} \\
		& ERGAS$\downarrow$  & 544.37   & 58.60  & 156.67         & 61.55        & 103.90 & 82.73   & 128.49  & 76.36  & 57.91  & {\ul 55.64}  & \textbf{43.63}  \\
		& SAM$\downarrow$   & 1.3764   & 0.0686 & 0.1546         & {\ul 0.0665} & 0.1425 & 0.0927  & 0.1125   & 0.0890 & 0.0691 & 0.0722       & \textbf{0.0576} \\ \hline
	\end{tabular}
}
\end{table*}

\begin{table*}[!t]
	\renewcommand{\arraystretch}{1.3}
	\caption{Metrics for HSI inpainting on the PC dataset. The best and the second best results are highlighted in \textbf{bold} and {\ul underline}, respectively.}
	\label{metrics for the Pavia HSIs dataset}
	\centering
	\resizebox{0.9\textwidth}{!}{
	\begin{tabular}{ccccccccccccc}
		\hline
		
		SR                   & Metrics & Observed & LRTFR & SNN            & KBR          & TNN    & SPCTV   & TNNTV   & FCTNTC & FCTNFR & TCTV         & HNN             \\ \hline
		\multirow{4}{*}{7\%} & PSNR$\uparrow$   & 14.68    & 30.05  & 21.72          & {\ul 36.95}  & 31.17       & 30.34  & 22.44  & 28.94  & 33.56        & 32.79  & \textbf{38.37}  \\
		& SSIM$\uparrow$   & 0.0380   & 0.8696 & 0.3756         & {\ul 0.9678} & 0.8822      & 0.8749 & 0.3975 & 0.8165 & 0.9288       & 0.9228 & \textbf{0.9799} \\
		& ERGAS$\downarrow$  & 722.79   & 114.90 & 319.96         & {\ul 56.13}  & 118.83      & 119.25 & 294.82 & 139.44 & 83.19        & 92.52  & \textbf{49.24}  \\
		& SAM$\downarrow$   & 1.3109   & 0.1311 & 0.1961         & {\ul 0.0806} & 0.1845      & 0.1369 & 0.1757 & 0.1742 & 0.1213       & 0.1513 & \textbf{0.0559} \\ \hline
		\multirow{4}{*}{5\%} & PSNR$\uparrow$   & 14.59    & 29.80  & 21.25          & 31.38        & 28.69       & 27.72  & 21.89  & 28.60  & {\ul 32.53}  & 30.86  & \textbf{35.04}  \\
		& SSIM$\uparrow$   & 0.0287   & 0.8629 & 0.3399         & 0.8912       & 0.8157      & 0.7863 & 0.3529 & 0.8041 & {\ul 0.9114} & 0.8861 & \textbf{0.9594} \\
		& ERGAS$\downarrow$  & 730.54   & 118.35 & 337.91         & 105.51       & 152.20      & 160.69 & 313.87 & 144.82 & {\ul 92.98}  & 113.63 & \textbf{72.15}  \\
		& SAM$\downarrow$   & 1.3557   & 0.1355 & 0.1931         & {\ul 0.1141} & 0.2003      & 0.1556 & 0.1703 & 0.1777 & 0.1272       & 0.1763 & \textbf{0.0717} \\ \hline
		\multirow{4}{*}{4\%} & PSNR$\uparrow$   & 14.54    & 29.53  & 21.04          & 28.58        & 26.91       & 25.97  & 21.67  & 28.31  & {\ul 31.68}  & 29.75  & \textbf{32.10}  \\
		& SSIM$\uparrow$   & 0.0240   & 0.8566 & 0.3255         & 0.8091       & 0.7446      & 0.7007 & 0.3344 & 0.7929 & {\ul 0.8947} & 0.8578 & \textbf{0.9225} \\
		& ERGAS$\downarrow$  & 734.34   & 122.08 & 346.05         & 145.46       & 183.00      & 196.45 & 322.02 & 149.69 & {\ul 102.09} & 128.31 & \textbf{103.87} \\
		& SAM$\downarrow$   & 1.3823   & 0.1421 & 0.1896         & {\ul 0.1295} & 0.2039      & 0.1656 & 0.1681 & 0.1831 & 0.1358       & 0.1911 & \textbf{0.0952} \\ \hline
	\end{tabular}
}
\end{table*}

\subsection{Experiments on HSI denoising}
This section carries out HSI denoising experiments to assess the performance of HNN based RPCA.
The compared methods include 
non-local meets global (NGMeet) \cite{NGMeet}, 
hyper-Laplacian regularized unidirectional low-rank tensor recovery (LLRT) \cite{LLRT}, 
tensor dictionary learning (TDL) \cite{TDL}, 
CTV \cite{CTV}, 
three-directional TNN (3DTNN) \cite{TGRSzheng2020}, 
3D log-based TNN (3DLogTNN) \cite{TGRSzheng2020}, 
and 
weighted non-local low-rank model with adaptive TV regularization (WNLRATV) \cite{WNLRATV}. 
Besides them, it is also compared with a deep learning based method, representative coefficient image with a learnable denoiser (RCILD) \cite{RCILD}.
The experiments simulate six scenarios:
\begin{enumerate}
	\item Case 1: I.i.d. Gaussian noise with $\sigma = 75$;
	\item Case 2: Non-i.i.d. Gaussian noise with band-varying standard deviations $\sigma \in [30, 100]$;
	\item Case 3: Building on Case 2, 1/3 of the bands are corrupted by impulse noise with varying ratios $p \in [5\%, 20\%]$;
	\item Case 4: Building on Case 2, 1/3 of the bands are corrupted by stripe noise with varying ratios $p \in [5\%, 20\%]$;
	\item Case 5: Building on Case 2, 1/3 of the bands are corrupted by deadline noise with varying ratios $p \in [5\%, 20\%]$;
	\item Case 6: Building on Case 2, the data is corrupted by a mixture of impulse, stripe, and deadline noise as described in the preceding cases.
\end{enumerate}

Tables \ref{tab:BA_denoisnig} and \ref{tab:PC_denoising} list the HSI denoising metrics on the Bay Area (BA) and Pavia Center (PC) datasets\footnote{[Online]. Avaialble: http://sipi.usc.edu/database/database.php}, respectively. The dimensions of the BA and PC datasets are cropped to $256 \times 256 \times 193$ and $200 \times 200 \times 80$, respectively, to facilitate efficient processing.
The noise intensity in this experiment configuration is severe, but HNN still exhibits robust performance. Notably, it surpasses CTV in terms of PSNR, with an improvement of 1.74 dB on the BA datasets for case 6. Visual inspection of the results in Fig. \ref{fig:pc_case6} reveals that methods including TDL, LTDL, CTV, 3DTNN and 3DLogTNN exhibit noticeable noise. Although NGMeet and WNLRATV effectively remove noise, they suffer from spectral distortion, due to the color mismatch. HNN stands out by generating an image that closely aligns with the ground truth.

\subsection{Experiments on HSI inpainting}
This subsection focuses on HSI inpainting, and we conduct experiments using the Bay Area (BA) and Pavia Center (PC) datasets. 

We compare the HNN method with ten methods, including nine model-based methods
(SNN \cite{liu2012tensor}, 
TNN \cite{lu2018exact}, 
TCTV \cite{TCTV}, 
Kronecker-Basis-Representation (KBR) \cite{KBR}, 
TNNTV \cite{TNNTV}, 
FCTN based tensor completition (FCTNTC) \cite{FCTN}, 
FCTN with factor regularization (FCTNFR) \cite{zhengFCTNFR2022}, 
and 
SPCTV \cite{SPCTV}) and a deep learning-based method (LRTFR\cite{LRTFR}), 
by varying the sampling rates (SR) at 4\%, 5\%, and 7\%. Tables \ref{metrics for the BayArea HSIs dataset} and \ref{metrics for the Pavia HSIs dataset} present the performance metrics on the BA and PC datasets, respectively.

\begin{figure*}[!t]
	\centering
	\includegraphics[width=0.9\linewidth]{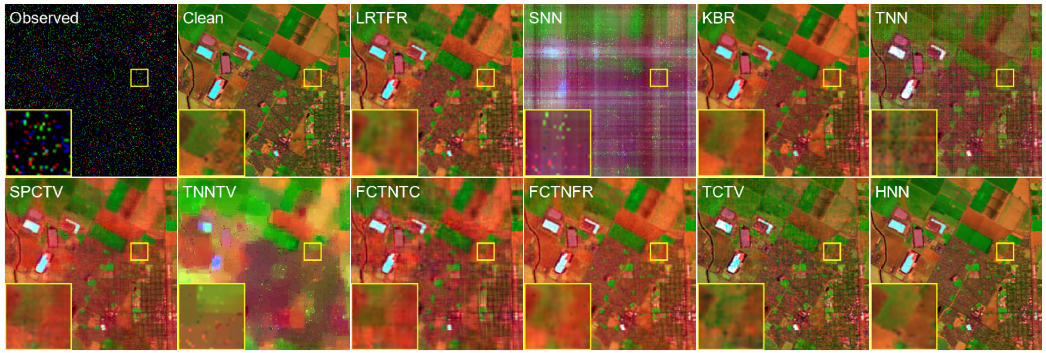}
	\caption{False-color images (band: 193-97-19) of all compared methods on the BA dataset.}
	\label{comparison_methods_BayArea_HSIs_dataset}
\end{figure*}

Table \ref{metrics for the BayArea HSIs dataset} shows that the HNN method outperforms other advanced methods in all evaluation metrics on the BA dataset. In terms of computational efficiency, the running times for the less effective KBR and TCTV methods are 1310.40s and 2020.76s, respectively, which are significantly higher than the HNN method's running time of 177.13s. This indicates that the HNN method not only achieves best performance but also offers improved efficiency. As depicted in Table \ref{metrics for the Pavia HSIs dataset}, the HNN approach consistently outperforms other methods on the PC dataset. For example, at a SR of 7\%, it is evident that HNN shows a superior PSNR value compared to KBR and FCTBFR by 1.42 dB and 4.81 dB, respectively. 

Fig. \ref{comparison_methods_BayArea_HSIs_dataset} depicts the false-color images on the BA dataset with a 5\% SR. A notable observation is that the SNN method simply filled in the missing areas, while the TNN, SPCTV, FCTNTC, and FCTCFR methods introduced grid-like stripes. Only the KBR, TCTV, and HNN successfully restored the HSIs in a visually appealing manner. Among these, the HNN method stands out with its good results, for example, sharper textures and colors that closely resemble the original clear images. This confirms the effectiveness of the HNN method in the task of HSI inpainting.
\begin{figure*}[]
	\centering
	\includegraphics[width=.9\linewidth]{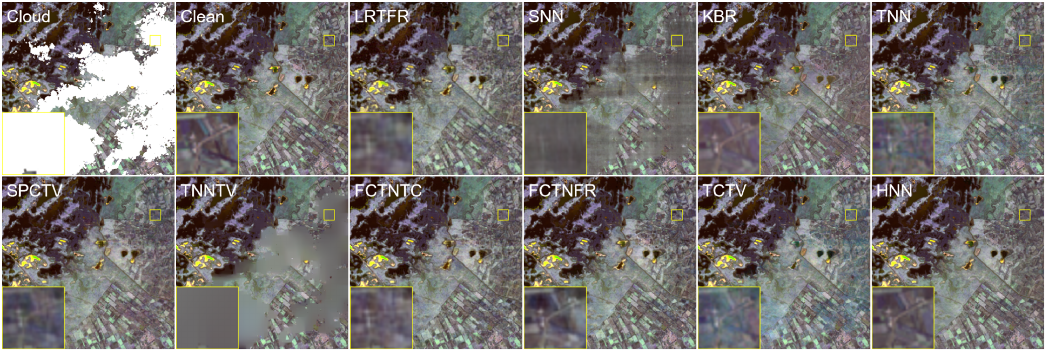}
	\caption{The false-color decloud images (the 5th time node, band 1-4-7) of all compared methods on the Jizzakh mountain dataset.}
	\label{fig_sim4}
\end{figure*}
\begin{table*}[!t]
	\caption{ Metrics on the Jizzakh mountain dataset. The best and the second best results are highlighted in \textbf{bold} and {\ul underline}, respectively.}
	\label{table1}
	\centering
	\resizebox{0.95\textwidth}{!}{
	\begin{tabular}{ccccccccccccc}
		\hline
		Time mode            & Metrics & Obs     & LRTFR & SNN    & KBR          & TNN    & SPCTV       & TNNTV        & FCTNTC & FCTNFR       & TCTV            & HNN             \\ \hline
		\multirow{4}{*}{\#1} & PSNR$\uparrow$   & 12.39   & 23.70   & 23.09  & 23.46        & 27.20  & 26.79       & 22.60        & 26.84  & 27.38        & {\ul 28.16}     & \textbf{28.24}  \\
		& SSIM$\uparrow$   & 0.5187  & 0.6809 & 0.7961 & 0.8726       & 0.8875 & 0.8717      & 0.7746       & 0.8753 & 0.8916       & {\ul 0.9089}    & \textbf{0.9323} \\
		& ERGAS$\downarrow$  & 764.33  & 157.65 & 182.84 & 173.29       & 113.07 & 117.77      & 191.43       & 117.39 & 110.69       & {\ul 101.99}    & \textbf{100.84} \\
		& SAM$\downarrow$   & 0.6122  & 0.1051 & 0.0751 & 0.0594       & 0.0593 & 0.0634      & 0.0673       & 0.0619 & 0.0603       & {\ul 0.0554}    & \textbf{0.0343} \\ \hline
		\multirow{4}{*}{\#2} & PSNR$\uparrow$   & 12.28   & 25.02  & 24.96  & 29.68        & 29.23  & 30.44       & 24.47        & 29.54  & {\ul 32.09}  & 30.73           & \textbf{32.48}  \\
		& SSIM$\uparrow$   & 0.5531  & 0.7198 & 0.8299 & 0.9472       & 0.9141 & 0.9176      & 0.8009       & 0.9054 & {\ul 0.9505} & 0.9386          & \textbf{0.9659} \\
		& ERGAS$\downarrow$  & 801.33  & 140.35 & 152.17 & 89.77        & 92.49  & 80.35       & 160.63       & 88.65  & {\ul 66.09}  & 78.44           & \textbf{64.46}  \\
		& SAM$\downarrow$   & 0.5067  & 0.1085 & 0.0420 & {\ul 0.0213} & 0.0332 & 0.0406      & 0.0403       & 0.0335 & 0.0245       & 0.0294          & \textbf{0.0156} \\ \hline
		\multirow{4}{*}{\#3} & PSNR$\uparrow$   & 14.01   & 25.71  & 27.60  & 33.81        & 32.49  & 33.40       & 27.26        & 32.22  & {\ul 35.21}  & 34.13           & \textbf{36.23}  \\
		& SSIM$\uparrow$   & 0.6819  & 0.7428 & 0.9017 & {\ul 0.9749} & 0.9511 & 0.9534      & 0.8892       & 0.9418 & 0.9739       & 0.9675          & \textbf{0.9859} \\
		& ERGAS$\downarrow$  & 594.71  & 128.28 & 110.81 & 55.57        & 63.09  & 56.60       & 114.86       & 64.70  & {\ul 46.07}  & 52.40           & \textbf{41.85}  \\
		& SAM$\downarrow$   & 0.3363  & 0.1020  & 0.0265 & 0.0114       & 0.0207 & 0.0215      & 0.0231       & 0.0255 & {\ul 0.0167} & 0.0180          & \textbf{0.0080} \\ \hline
		\multirow{4}{*}{\#4} & PSNR$\uparrow$   & 13.24   & 25.32  & 26.34  & 31.83        & 30.02  & 32.01       & 25.64        & 30.83  & {\ul 33.28}  & 31.81           & \textbf{34.24}  \\
		& SSIM$\uparrow$   & 0.7193  & 0.7358 & 0.8857 & {\ul 0.9726} & 0.9374 & 0.9506      & 0.8613       & 0.9369 & 0.9697       & 0.9571          & \textbf{0.9802} \\
		& ERGAS$\downarrow$  & 663.22  & 133.29 & 128.05 & 68.36        & 82.72  & 65.55       & 138.33       & 75.18  & {\ul 56.89}  & 67.37           & \textbf{51.29}  \\
		& SAM$\downarrow$   & 0.3519  & 0.0988 & 0.0245 & 0.0095       & 0.0206 & 0.0158      & 0.0240       & 0.0190 & 0.0137       & 0.0179          & \textbf{0.0082} \\ \hline
		\multirow{4}{*}{\#5} & PSNR$\uparrow$   & 10.54   & 25.67  & 23.02  & 27.59        & 28.58  & 30.26       & 23.85        & 29.28  & {\ul 31.72}  & 29.42           & \textbf{32.59}  \\
		& SSIM$\uparrow$   & 0.4817  & 0.7321 & 0.7669 & 0.9340       & 0.8807 & 0.8992      & 0.7331       & 0.8765 & {\ul 0.9386} & 0.9146          & \textbf{0.9610} \\ 
		& ERGAS$\downarrow$  & 1052.98 & 124.84 & 185.63 & 108.97       & 95.38  & 78.63       & 165.94       & 87.76  & {\ul 66.56}  & 87.25           & \textbf{61.15}  \\
		& SAM$\downarrow$   & 0.6842  & 0.0853 & 0.0475 & {\ul 0.0238} & 0.0386 & 0.0297      & 0.0448       & 0.0346 & 0.0259       & 0.0370          & \textbf{0.0159} \\ \hline
		\multirow{4}{*}{\#6} & PSNR$\uparrow$   & 12.39   & 25.91  & 24.03  & 24.99        & 28.15  & 31.16       & 23.45        & 30.42  & {\ul 31.95}  & 27.90           & \textbf{32.80}  \\
		& SSIM$\uparrow$   & 0.6147  & 0.7253 & 0.8421 & 0.9229       & 0.9025 & 0.9247      & 0.8154       & 0.9120 & {\ul 0.9415} & 0.9153          & \textbf{0.9660} \\
		& ERGAS$\downarrow$  & 699.76  & 116.97 & 159.04 & 140.87       & 98.41  & 68.07       & 169.59       & 74.07  & {\ul 62.35}  & 100.54          & \textbf{58.96}  \\
		& SAM$\downarrow$   & 0.4981  & 0.0709 & 0.0535 & 0.0362       & 0.0438 & 0.0312      & 0.0551       & 0.0343 & {\ul 0.0279} & 0.0385          & \textbf{0.0187} \\ \hline
		\multirow{4}{*}{\#7} & PSNR$\uparrow$   & 20.65   & 23.97  & 31.62  & 31.33        & 32.98  & 32.83       & 31.61        & 32.18  & 30.86        & \textbf{33.82}  & {\ul 33.21}     \\
		& SSIM$\uparrow$   & 0.9162  & 0.6558 & 0.9647 & 0.9709       & 0.9733 & 0.9655      & 0.9638       & 0.9646 & 0.9632       & \textbf{0.9774} & {\ul 0.9756}    \\
		& ERGAS$\downarrow$  & 248.40  & 154.24 & 68.02  & 73.98        & 59.48  & 60.15       & 68.41        & 63.95  & 75.15        & \textbf{54.14}  & {\ul 57.84}     \\
		& SAM$\downarrow$   & 0.0956  & 0.1131 & 0.0095 & 0.0098       & 0.0099 & 0.0121      & {\ul 0.0082} & 0.0122 & 0.0125       & 0.0092          & \textbf{0.0081} \\ \hline
		\multirow{4}{*}{\#8} & PSNR$\uparrow$   & 14.17   & 23.89  & 25.66  & 26.79        & 29.07  & {\ul 29.97} & 25.19        & 29.29  & 29.93        & 29.94           & \textbf{31.07}  \\
		& SSIM$\uparrow$   & 0.7535  & 0.6841 & 0.8883 & 0.9289       & 0.9320 & 0.9268      & 0.8765       & 0.9247 & 0.9411       & {\ul 0.9437}    & \textbf{0.9613} \\
		& ERGAS$\downarrow$  & 541.15  & 150.68 & 130.14 & 119.73       & 89.92  & {\ul 79.98} & 137.46       & 85.99  & 80.78        & 82.59           & \textbf{71.08}  \\
		& SAM$\downarrow$   & 0.2950  & 0.1306 & 0.0352 & {\ul 0.0296} & 0.0304 & 0.0563      & 0.0311       & 0.0370 & 0.0377       & 0.0299          & \textbf{0.0199} \\ \hline
		\multirow{4}{*}{\#9} & PSNR$\uparrow$   & 14.61   & 24.05  & 26.19  & 27.66        & 29.64  & 30.50       & 25.47        & 29.79  & 30.08        & {\ul 30.91}     & \textbf{31.47}  \\
		& SSIM$\uparrow$   & 0.7734  & 0.6945 & 0.9105 & 0.9439       & 0.9426 & 0.9369      & 0.9024       & 0.9353 & 0.9488       & {\ul 0.9560}    & \textbf{0.9664} \\
		& ERGAS$\downarrow$  & 514.77  & 149.08 & 124.67 & 108.15       & 83.85  & 75.66       & 135.52       & 81.86  & 79.63        & {\ul 72.40}     & \textbf{68.25}  \\
		& SAM$\downarrow$   & 0.2575  & 0.1172 & 0.0261 & {\ul 0.0201} & 0.0227 & 0.0363      & 0.0236       & 0.0287 & 0.0290       & 0.0218          & \textbf{0.0140} \\ \hline
		\multirow{4}{*}{Mean} & PSNR$\uparrow$ & 13.81 & 24.81 & 25.83 & 28.57 & 29.71 & 30.82 & 25.51 & 30.04 & {\ul 31.39} & 30.76 & \textbf{32.48} \\
		& SSIM$\uparrow$   & 0.6681  & 0.7079 & 0.8651 & 0.9409       & 0.9246 & 0.9274      & 0.8463       & 0.9192 & {\ul 0.9465} & 0.9421          & \textbf{0.9661} \\
		& ERGAS$\downarrow$  & 686.20  & 140.14 & 142.25 & 110.09       & 87.94  & 77.73       & 146.45       & 83.56  & {\ul 73.64}  & 79.27           & \textbf{65.81}  \\
		& SAM$\downarrow$   & 0.5160  & 0.1278 & 0.1333 & 0.1047       & 0.0853 & 0.0766      & 0.1300       & 0.0787 & {\ul 0.0684} & 0.0766          & \textbf{0.0550} \\ \hline
	\end{tabular}
}
\end{table*}

\begin{table*}[!t]
	\caption{Metrics on the Jizzakh urban dataset. The best and the second best results are highlighted in \textbf{bold} and {\ul underline}, respectively.}
	\label{table2}
	\centering
	\resizebox{0.95\textwidth}{!}{
	\begin{tabular}{ccccccccccccc}
		\hline
		Time mode            & Metrics & Obs     & LRTFR & SNN    & KBR          & TNN    & SPCTV       & TNNTV        & FCTNTC & FCTNFR       & TCTV            & HNN             \\ \hline
		\multirow{4}{*}{\#1}  & PSNR$\uparrow$  & 10.71  & 21.71  & 24.92  & \textbf{26.41}  & 24.69  & 25.16        & 25.03  & 24.84  & 24.50        & 25.32  & {\ul 26.30}     \\
		& SSIM$\uparrow$  & 0.4982 & 0.5649 & 0.7665 & {\ul 0.8487}    & 0.7846 & 0.7877       & 0.7506 & 0.7880 & 0.8070       & 0.8174 & \textbf{0.8515} \\
		& ERGAS$\downarrow$ & 892.96 & 184.14 & 137.10 & \textbf{114.27} & 140.64 & 131.42       & 135.40 & 136.45 & 141.58       & 131.03 & {\ul 116.67}    \\
		& SAM$\downarrow$  & 0.6135 & 0.0826 & 0.0403 & {\ul 0.0365}    & 0.0457 & 0.0400       & 0.0401 & 0.0441 & 0.0449       & 0.0417 & \textbf{0.0311} \\  \hline
		\multirow{4}{*}{\#2}  & PSNR$\uparrow$  & 12.31  & 23.31  & 24.39  & 26.89           & 25.56  & 27.43        & 24.38  & 26.95  & {\ul 27.73}  & 27.15  & \textbf{28.76}  \\
		& SSIM$\uparrow$  & 0.5487 & 0.6555 & 0.8036 & {\ul 0.9062}    & 0.8408 & 0.8737       & 0.7844 & 0.8650 & 0.8985       & 0.8806 & \textbf{0.9279} \\
		& ERGAS$\downarrow$ & 753.35 & 163.37 & 156.03 & 115.24          & 134.02 & 108.08       & 156.19 & 114.02 & {\ul 104.07} & 112.21 & \textbf{93.24}  \\
		& SAM$\downarrow$  & 0.5071 & 0.0831 & 0.0463 & 0.0356          & 0.0475 & 0.0315       & 0.0463 & 0.0353 & {\ul 0.0308} & 0.0410 & \textbf{0.0215} \\ \hline
		\multirow{4}{*}{\#3}  & PSNR$\uparrow$  & 14.08  & 24.96  & 26.91  & 30.99           & 28.56  & 30.90        & 26.89  & 30.43  & {\ul 31.47}  & 30.86  & \textbf{32.88}  \\
		& SSIM$\uparrow$  & 0.6720 & 0.6977 & 0.8837 & {\ul 0.9589}    & 0.9109 & 0.9322       & 0.8686 & 0.9234 & 0.9474       & 0.9387 & \textbf{0.9680} \\
		& ERGAS$\downarrow$ & 566.35 & 134.07 & 115.47 & 71.75           & 94.38  & 71.86        & 116.48 & 75.89  & {\ul 67.32}  & 73.07  & \textbf{57.39}  \\
		& SAM$\downarrow$  & 0.3369 & 0.0712 & 0.0265 & {\ul 0.0162}    & 0.0271 & 0.0180       & 0.0264 & 0.0203 & 0.0178       & 0.0224 & \textbf{0.0109} \\ \hline
		\multirow{4}{*}{\#4}  & PSNR$\uparrow$  & 14.11  & 24.66  & 26.79  & 29.53           & 27.63  & 30.22        & 26.59  & 29.79  & {\ul 30.64}  & 29.80  & \textbf{32.63}  \\
		& SSIM$\uparrow$  & 0.7192 & 0.6824 & 0.8707 & {\ul 0.9498}    & 0.8906 & 0.9242       & 0.8561 & 0.9192 & 0.9426       & 0.9241 & \textbf{0.9652} \\
		& ERGAS$\downarrow$ & 559.53 & 137.50 & 115.78 & 85.12           & 104.00 & 77.00        & 119.17 & 80.97  & {\ul 73.42}  & 81.27  & \textbf{58.71}  \\
		& SAM$\downarrow$  & 0.3519 & 0.0703 & 0.0283 & 0.0217          & 0.0304 & 0.0193       & 0.0282 & 0.0217 & {\ul 0.0192} & 0.0256 & \textbf{0.0128} \\ \hline
		\multirow{4}{*}{\#5}  & PSNR$\uparrow$  & 10.51  & 24.70  & 23.69  & 28.00           & 25.48  & 28.28        & 23.86  & 27.83  & {\ul 28.72}  & 26.87  & \textbf{29.91}  \\
		& SSIM$\uparrow$  & 0.4747 & 0.6715 & 0.7441 & {\ul 0.9092}    & 0.8088 & 0.8617       & 0.7160 & 0.8485 & 0.8976       & 0.8619 & \textbf{0.9339} \\
		& ERGAS$\downarrow$ & 965.49 & 133.15 & 163.06 & 97.45           & 130.22 & 94.18        & 159.82 & 98.63  & {\ul 89.64}  & 110.73 & \textbf{79.05}  \\
		& SAM$\downarrow$  & 0.6844 & 0.0611 & 0.0439 & 0.0303          & 0.0494 & {\ul 0.0299} & 0.0458 & 0.0340 & 0.0311       & 0.0456 & \textbf{0.0201} \\ \hline
		\multirow{4}{*}{\#6}  & PSNR$\uparrow$  & 11.22  & 24.90  & 25.84  & {\ul 29.78}     & 24.83  & 29.17        & 25.75  & 28.52  & 28.32        & 25.17  & \textbf{30.64}  \\
		& SSIM$\uparrow$  & 0.6041 & 0.6507 & 0.8245 & {\ul 0.9271}    & 0.8323 & 0.8877       & 0.8073 & 0.8747 & 0.8961       & 0.8562 & \textbf{0.9378} \\
		& ERGAS$\downarrow$ & 772.65 & 125.15 & 120.82 & {\ul 77.19}     & 137.24 & 81.76        & 121.98 & 88.08  & 90.01        & 133.95 & \textbf{69.32}  \\
		& SAM$\downarrow$  & 0.4985 & 0.0539 & 0.0246 & {\ul 0.0148}    & 0.0394 & 0.0197       & 0.0267 & 0.0240 & 0.0242       & 0.0378 & \textbf{0.0142} \\ \hline
		\multirow{4}{*}{\#7}  & PSNR$\uparrow$  & 18.74  & 24.19  & 30.30  & {\ul 32.69}     & 29.02  & 31.97        & 30.19  & 30.21  & 29.98        & 29.64  & \textbf{33.05}  \\
		& SSIM$\uparrow$  & 0.9108 & 0.6890 & 0.9581 & \textbf{0.9723} & 0.9502 & 0.9631       & 0.9595 & 0.9573 & 0.9588       & 0.9568 & {\ul 0.9721}    \\
		& ERGAS$\downarrow$ & 292.35 & 144.03 & 78.54  & {\ul 60.25}     & 90.79  & 64.51        & 79.57  & 79.13  & 80.73        & 84.09  & \textbf{57.42}  \\
		& SAM$\downarrow$  & 0.0958 & 0.0758 & 0.0114 & 0.0083          & 0.0128 & {\ul 0.0083} & 0.0111 & 0.0104 & 0.0100       & 0.0117 & \textbf{0.0073} \\ \hline
		\multirow{4}{*}{\#8}  & PSNR$\uparrow$  & 13.42  & 22.06  & 26.36  & {\ul 28.07}     & 25.59  & 27.53        & 26.45  & 26.74  & 26.69        & 26.26  & \textbf{28.11}  \\
		& SSIM$\uparrow$  & 0.7359 & 0.6159 & 0.8743 & {\ul 0.9181}    & 0.8768 & 0.8935       & 0.8674 & 0.8883 & 0.9040       & 0.8933 & \textbf{0.9203} \\
		& ERGAS$\downarrow$ & 553.45 & 173.80 & 114.75 & \textbf{92.56}  & 126.79 & 98.76        & 114.00 & 108.27 & 108.49       & 118.66 & {\ul 92.74}     \\
		& SAM$\downarrow$  & 0.2967 & 0.0765 & 0.0211 & {\ul 0.0162}    & 0.0257 & 0.0186       & 0.0213 & 0.0211 & 0.0217       & 0.0245 & \textbf{0.0155} \\ \hline
		\multirow{4}{*}{\#9}  & PSNR$\uparrow$  & 14.50  & 22.32  & 26.48  & \textbf{28.45}  & 26.67  & 27.83        & 26.56  & 26.82  & 26.68        & 27.82  & {\ul 28.21}     \\
		& SSIM$\uparrow$  & 0.7625 & 0.6427 & 0.8965 & \textbf{0.9333} & 0.9028 & 0.9109       & 0.8921 & 0.9055 & 0.9173       & 0.9218 & {\ul 0.9329}    \\
		& ERGAS$\downarrow$ & 502.23 & 175.25 & 117.70 & \textbf{92.13}  & 114.61 & 99.04        & 117.36 & 111.52 & 113.68       & 100.09 & {\ul 95.54}     \\
		& SAM$\downarrow$  & 0.2585 & 0.0804 & 0.0213 & {\ul 0.0163}    & 0.0235 & 0.0176       & 0.0213 & 0.0204 & 0.0217       & 0.0211 & \textbf{0.0145} \\ \hline
		\multirow{4}{*}{Mean} & PSNR$\uparrow$  & 13.29  & 23.64  & 26.19  & {\ul 28.98}     & 26.45  & 28.72        & 26.19  & 28.02  & 28.30        & 27.65  & \textbf{30.05}  \\
		& SSIM$\uparrow$  & 0.6584 & 0.6523 & 0.8469 & {\ul 0.9248}    & 0.8664 & 0.8927       & 0.8335 & 0.8855 & 0.9077       & 0.8945 & \textbf{0.9344} \\
		& ERGAS$\downarrow$ & 680.84 & 153.67 & 126.60 & {\ul 91.21}     & 120.51 & 93.88        & 126.50 & 101.02 & 98.96        & 107.02 & \textbf{82.41}  \\
		& SAM$\downarrow$  & 0.5059 & 0.1104 & 0.0861 & {\ul 0.0641}    & 0.0884 & 0.0673       & 0.0871 & 0.0722 & 0.0702       & 0.0791 & \textbf{0.0537} \\  \hline
	\end{tabular}
}
\end{table*}
\begin{figure}[!t]
	\centering
	\includegraphics[width=.85\linewidth]{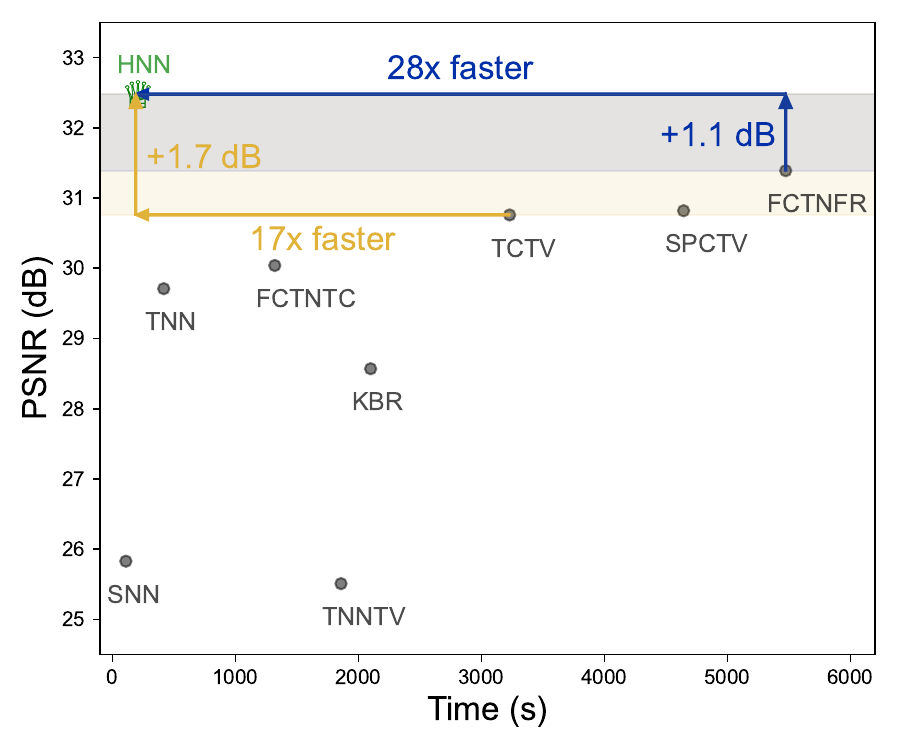}
	\caption{The scatter plot of PSNR versus time for the Jizzakh mountain dataset.}
	\label{fig_sim8}
\end{figure}

\subsection{Experiment on multi-temporal cloud removal}
In this section, we concentrate on a specialized inpainting task: multi-temporal cloud removal. In this scenario, areas covered by clouds are considered missing pixels, and the objective is to reconstruct these pixels. However, clouds have distinctive shape patterns, making this task more challenging than randomly pixel missing cases.

A series of comprehensive synthetic experiments are conducted to evaluate the effectiveness of the HNN method. The synthetic datasets comprise two remote sensing images captured over Jizzakh by the Landsat-8 satellite, encompassing mountain and urban areas, with seven channels and nine time nodes. The size of these data is $512 \times 512 \times 7 \times 9$. 
To ensure the fidelity of the synthetic data to real-world conditions, we selected nine real cloud masks with varying levels of cloud coverage from the remote sensing cloud detection dataset.

Table \ref{table1} summarizes the performance metrics of all the compared methods on the Jizzakh mountain dataset. HNN demonstrates the most favorable performance. It consistently achieves the highest PSNR and SSIM scores while maintaining the lowest ERGAS and SAM scores across nearly all time nodes.

Table \ref{table2} presents the performance metrics on the Jizzakh urban dataset. As with the results obtained from the Jizzakh mountain dataset, it is evident that the HNN consistently achieves the best performance metrics across the majority of time nodes, with the second-best results observed at time nodes 1 and 9. In comparison to other methods, the results of HNN demonstrate both stability and superiority.

To further highlight the superiority of the HNN method, Fig. \ref{fig_sim4} displays 
false-color declouded images. To facilitate a detailed comparison, specific regions within each image are magnified for enhanced visibility. In Fig. \ref{fig_sim4}, the SNN, SPCTV, and FCTNTC methods are observed to fill the cloud regions rather haphazardly, while the TNN and TCTV methods do not successfully reconstruct the original texture details beneath the clouds. The KBR and FCTNFR methods exhibit color discrepancies compared to the original images. Conversely, the HNN method effectively restores the intricate texture details, resulting in the restored images that closely resemble the original cloud-free images.

\subsection{Disccusion}

\subsubsection{Running Time}


Table \ref{tab:running_time} compares the execution times of all methods. A GPU-compatible implementation of the HNN method is also provided. For inpainting tasks, the GPU-based HNN method achieves both the fastest runtime and highest performance among all methods compared. Fig. \ref{fig_sim8} presents a scatter plot of the Jizzakh mountain dataset, clearly illustrating the performance-time relationship. SNN is the only method faster than GPU-based HNN for inpainting tasks, but SNN does not yield satisfactory restoration results. Remarkably, HNN achieves the highest PSNR while maintaining relatively short execution times. In contrast, although TCTV and FCTNFR yield impressive results, their lengthy execution times are a concern. Similarly, for denoising tasks, while TDL and RCILD have shorter execution times, their PSNR do not meet expectations.

\begin{table*}[htbp]
	\centering
	\caption{The running time of compared methods (in seconds).}
	\resizebox{0.95\textwidth}{!}{
\begin{tabular}{ccccccccccccc}
\hline
Task                            & Datasets & LRTFR   & SNN            & KBR      & TNN    & SPCTV   & TNNTV   & FCTNTC   & FCTNFR  & TCTV    & HNN    & HNN(GPU)       \\ \hline
\multirow{2}{*}{HSI Inpainting} & BA       & 111.76  & {\ul 71.19}    & 1310.4   & 219.42 & 3272.72 & 1108.82 & 244.23   & 285.69  & 2020.76 & 147.99 & \textbf{71.11} \\
                                & PC       & 22.55   & {\ul 18.45}    & 276.63   & 39.38  & 885.58  & 211.36  & 68.22    & 101.13  & 275.12  & 40.53  & \textbf{15.66} \\ \hline
\multirow{2}{*}{Cloud Removal}  & Mountain & 111.69  & {\ul 109.54}   & 2099.33  & 417.33 & 4643.82 & 1858.60 & 1320.26  & 5474.12 & 3229.33 & 129.36 & \textbf{40.82} \\
                                & Urban    & 154.30  & {\ul 71.71}    & 1730.17  & 348.06 & 3953.04 & 1828.23 & 968.53   & 4682.67 & 3105.27 & 120.21 & \textbf{43.94} \\ \hline
Task                            & Datasets & LLRT    & TDL            & LTDL     & NGMeet & CTV     & 3DTNN   & 3DLogTNN & WNLRATV & RCILD   & HNN    & HNN(GPU)       \\ \hline
\multirow{2}{*}{HSI Denoising}  & BA       & 1456.47 & {\ul 32.73} & 16160.75 & 75.23  & 118.97  & 259.68  & 387.73   & 646.28  & \textbf{4.23}       & 84.85  & 40.58    \\
                                & PC       & 399.67  &{\ul 6.80}  & 15060.76 & 35.03  & 21.48   & 46.25   & 70.4     & 200.43  & \textbf{1.45}       & 19.53  &  9.36     \\ \hline
\end{tabular}
	}
	\label{tab:running_time}%
\end{table*}%

\subsubsection{Influences of Different Norms and Haar Wavelet Transform}

\begin{figure*}[]
\centering
	\includegraphics[width=.85\linewidth]{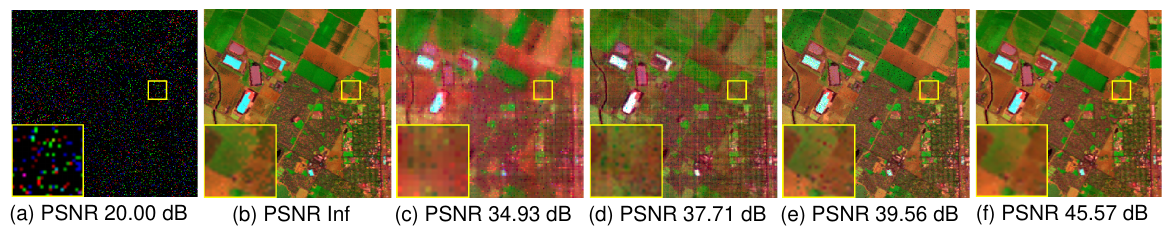}
	\caption{False-color images (band: 193-97-19) of all methods on the BA dataset. (a) Observed image. (b) Original image. (c) Sum nuclear norm + Haar wavelet transform. (d) Tensor nuclear norm + Haar wavelet transform. (e) Nuclear norm without Haar wavelet transform. (f) Nuclear norm + Haar wavelet transform (Our proposed).}
	\label{ablation experiment}
\end{figure*}
Choosing an appropriate norm is a critical factor in the performance of our HNN method. However, the Haar wavelet transform can be readily extended to different nuclear norms. Here, we conducted a comparative analysis integrating the Haar wavelet transform with the sum nuclear norm and the tensor nuclear norm for image inpainting. As illustrated in Fig. \ref{ablation experiment}(c), (d), and (f), the nuclear norm is a particularly favorable choice, demonstrating significant performance advantages over other norms. Furthermore, we compared the performance of the nuclear norm both with and without the Haar wavelet transform. Our proposed method, combining the nuclear norm with the Haar wavelet transform (HNN), significantly outperforms the method using only the nuclear norm without the Haar wavelet transform; see Fig.\ref{ablation experiment}(e) and (f). This illustrates that incorporating the Haar wavelet transform notably enhances the performance of image inpainting.

\subsubsection{Numerical Analysis of Convergence}

\begin{table}[!t]
\caption{ Metrics on the Jizzakh mountain dataset with varying temporal numbers.}
\label{table4}
\centering
\resizebox{0.95\linewidth}{!}{
 \begin{tabular}{cccccccc}
	 \hline
	 Metrics & 3      & 4      & 5      & 6      & 7      & 8      & 9      \\ \hline
	 PSNR    & 24.33  & 28.86  & 30.58  & 31.80  & 32.21  & 32.09  & 32.48  \\
	 SSIM    & 0.9142 & 0.9488 & 0.9527 & 0.9609 & 0.9645 & 0.9619 & 0.9661 \\
	 ERGAS   & 161.97 & 98.36  & 84.04  & 73.06  & 68.93  & 71.31  & 65.81  \\
	 SAM    & 0.0847 & 0.0606 & 0.0551 & 0.0539 & 0.0526 & 0.0578 & 0.0550 \\ \hline
	 \end{tabular}
 }
\end{table}

\begin{figure}[!t]
	\centering
	\includegraphics[width=.85\linewidth]{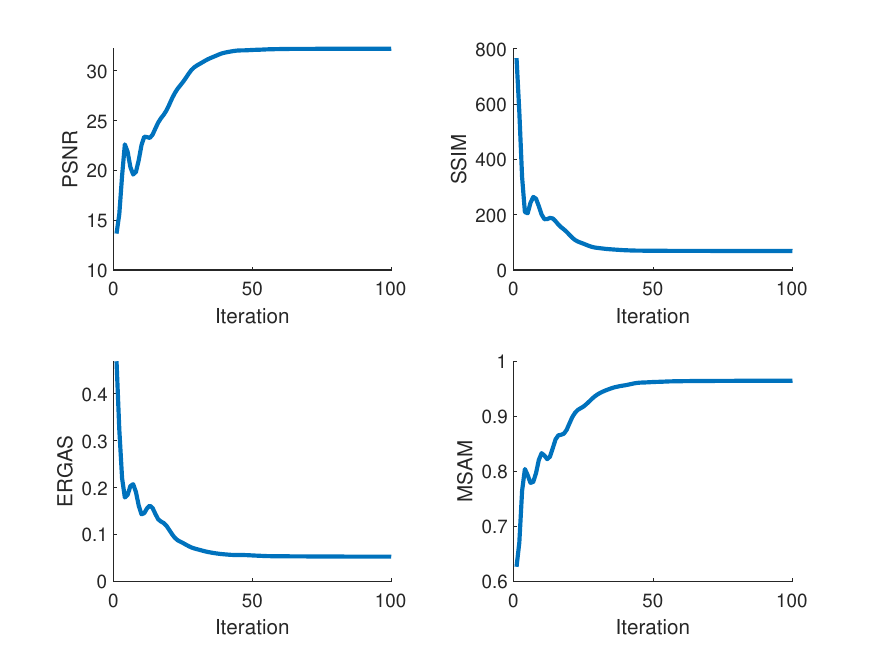}
	\caption{Metric curves versus iteration number.}
	\label{fig_sim9}
\end{figure}

The algorithm's convergence is guaranteed by the fact that the current model satisfies the sufficient conditions for ADMM convergence. To further validate its convergence, we conduct a numerical analysis of the HNN's convergence. Fig. \ref{fig_sim9} depicts the curves of PSNR, SSIM, ERGAS, and SAM values versus the number of iterations on the Jizzakh mountain dataset. Notably, after 40 iterations, their values reach a stable state. This observation provides further evidence for the convergence of the ADMM algorithm employed in this study for solving the HNN model.

\subsubsection{Influence of the Temporal Number}

This section explores the influence of temporal number on HNN's performance, by varying the temporal number on the Jizzakh mountain dataset. Recalling the results with 9 time nodes (listed in Table \ref{table1}), it is shown that, if not considering HNN, FCTNFC and TCTV are two best performer among the existing methods, achieving PSNR values of 31.39 dB and 30.76 dB, respectively. However, Table \ref{table4} shows that the HNN achieves PSNR values of 31.80 dB with only 6 time nodes, surpassing the performance of the existing methods by a considerable improvement. This result further confirms the effectiveness of the HNN method.

\section{Conclusion}
This paper introduces the HNN as a regularization term for remote sensing image restoration. HNN effectively combines low-rank and smoothness priors and is computationally efficient. Its superior performance in inpainting, denoising, and cloud removal tasks makes it a valuable tool for enhancing the quality and utility of remote sensing imagery. Future research can explore the application of HNN in other remote sensing image processing tasks, such as HSI unmixing and classification.




%





\ifCLASSOPTIONcaptionsoff
\newpage
\fi

\bibliographystyle{IEEEtran}

\bibliography{IEEEabrv}





\end{document}